\documentclass[12pt,amssymb,qsymbols]{article}
\pdfoutput=1
\usepackage{tabularx}
\usepackage[pdftex]{graphicx}
\usepackage{amsmath}
\usepackage{amssymb}
\usepackage{longtable}
\usepackage{verbatim}
\usepackage{hyperref}
\usepackage{color}

\makeatletter

\setlongtables

\newcommand{\msbar}{{\overline{\rm MS}}}

\newcommand{\bea}{\begin{eqnarray}}
\newcommand{\eea}{\end{eqnarray}}
\newcommand{\beq}{\begin{equation}}
\newcommand{\eeq}{\end{equation}}
\newcommand{\ec}{\end{center}}
\newcommand{\bc}{\begin{center}}
\newcommand{\gev}{{\rm GeV}}

\newcommand{\pdir}{p\kern -5.2pt\raise 0.2ex\hbox {/}}

\newcommand{\vdir}{v\kern -5.75pt\raise 0.15ex\hbox {/}}
\newcommand{\kdir}{k\kern -5.75pt\raise 0.15ex\hbox {/}}
\newcommand{\epsdir}{\epsilon\kern -5.0pt\raise 0.15ex\hbox {/}}
\newcommand{\bvdir}{\bar{v}\kern -5.75pt\raise 0.15ex\hbox {/}}
\newcommand{\Ddir}{D\kern -7.75pt\raise 0.20ex\hbox {/}}
\newcommand{\Adir}{A\kern -7.75pt\raise 0.20ex\hbox {/}}
\newcommand{\ldir}{l\kern -5.0pt\raise 0.2ex\hbox{/}}
\newcommand{\varepsdir}{\varepsilon\kern -5.5pt\raise 0.15ex\hbox{/}}

\def \eff{{\text{eff}}}

\newcommand{\qq}{(q^2)}

\newcommand{\nn}{\nonumber}
\makeatother

\definecolor{nicered}{rgb}{0.7,0.1,0.1}
\definecolor{nicegreen}{rgb}{0.1,0.5,0.1}
\hypersetup{colorlinks,citecolor= nicegreen,linkcolor= nicered}

\begin{document}
\thispagestyle{empty} 
\begin{flushright}
\begin{tabular}{l}
{\small \tt ICCUB-12-305}\\
{\small \tt LAL 12-182}\\
{\small \tt LPT 12-48}\\
\end{tabular}
\end{flushright}
\begin{center}
\vskip 2.5cm\par
{\par\centering \textbf{\LARGE  
\Large \bf Complementarity of the constraints on New Physics }}\\
\vskip .25cm\par
{\par\centering \textbf{\LARGE  
\Large  from $B_s\to \mu^+\mu^-$ and  from $B\to K \ell^+\ell^-$ decays  } }\\
\vskip 1.25cm\par
{\scalebox{.82}{\par\centering \large  
\sc Damir Be\v cirevi\'c$^a$, Nejc Ko\v snik$^b$, Federico Mescia$^c$, Elia Schneider$^d$ }}
{\par\centering \vskip 0.3 cm\par}
{\sl \small
$^a$~Laboratoire de Physique Th\'eorique (B\^at.~210)~\footnote{Laboratoire de Physique Th\'eorique est une unit\'e mixte de recherche du CNRS, UMR 8627.}\\
Universit\'e Paris Sud, Centre d'Orsay, F-91405 Orsay Cedex, France}\\
{\par\centering \vskip 0.3 cm\par}
{\sl \small
$^b$~Laboratoire de l'Acc\'el\'erateur Lin\'eaire, Centre d'Orsay, Universit\'e de Paris-Sud XI,\\
 B.P. 34,  B\^atiment 200, 91898 Orsay Cedex, France, and \\
J. Stefan Institute, Jamova 39, P. O. Box 3000, 1001 Ljubljana, Slovenia }\\
{\par\centering \vskip 0.3 cm\par}
{\sl \small
$^c$~Departament d'Estructura i Constituents de la Mat\`eria and
Institut de Ci\`encies del\\
Cosmos (ICCUB), Universitat de Barcelona, 08028 Barcelona, Spain}\\
{\par\centering \vskip 0.3 cm\par}
{\sl \small
$^d$~Dipartimento di Fisica, Universit\`a degli Studi di Trento,  and
INFN Gruppo collegato di Trento \\
Via Sommarive 14, Povo (Trento), I-38123 Italy.}\\

{\vskip 1.2cm\par}
\end{center}

\vskip 0.55cm
\begin{abstract}
We discuss the advantages of combining  the experimental bound on ${\rm Br}(B_s\to \mu^+\mu^-)$ and the measured ${\rm Br}(B\to K\ell^+\ell^-)$ to get the model independent constraints on physics beyond the Standard Model. Since the two decays give complementary information, one can study not only the absolute values of the Wilson coefficients that are zero in the Standard Model, but also their phases. To identify the sector in which the new physics might appear, information about the shapes of the transverse asymmetries in $B\to K^\ast \ell^+\ell^-$ at low $q^2$'s can be particularly useful. We also emphasize the importance of measuring the forward-backward asymmetry in $B\to K\ell^+\ell^-$ decay at large $q^2$'s. 
\end{abstract}
\newpage
\setcounter{page}{1}
\setcounter{footnote}{0}
\setcounter{equation}{0} 
\noindent

\renewcommand{\thefootnote}{\arabic{footnote}}

\setcounter{footnote}{0}
\section{Introduction and basic formulas}

Ever since the first observation of $B\to K^\ast \gamma$~\cite{Ammar:1993sh}, the decay modes governed by the loop-induced $b\to s$ transition play a major role in the search of signals of  physics beyond the Standard Model (BSM) in low energy experiments. After years of experimental and theoretical effort we now know that  the observed decay rates and several  conveniently defined observables are consistent with the Standard Model (SM)  predictions, thus leaving little room for New Physics (NP). Since the non-perturbative QCD (hadronic) uncertainties are still large in most cases, the comparison between theory and experiment is not yet at the precision level and the quantitative statements about the size of possible NP contributions are often subjects of controversies. 

Much of the experimental activity has been devoted to $B\to K^\ast \ell^+\ell^-$ decay, for which the improvement on theoretical (hadronic) uncertainties is hard to achieve, apart from a few asymmetries that will be studied at CMS, LHCb and in the new generation of $B$-physics experiments. In contrast to the decay to $K^\ast$, a substantial improvement in the determination of the hadronic form factors entering the theoretical description of $B\to K \ell^+\ell^-$ is realistic to expect very soon. In that respect the recently reported result on ${\rm Br}(B\to K \ell^+\ell^-)$ by BaBar~\cite{exp-BKll} is likely to become a major constraint in the NP searches. The ongoing experimental effort to  detect another $b\to s$ mediated decay, $B_s\to\mu^+\mu^-$, has been greatly improved after LHCb was able to set the upper bound on ${\rm Br}(B_s\to\mu^+\mu^-)$ that got very close to the value predicted in the SM~\cite{LHCb-Bmumu}.
In this paper we discuss how these two decay modes can be combined to give us complementary information about the potential physics contributions from BSM.

The most important effect of physics BSM in $B_s\to\mu^+\mu^-$  is expected to come from a coupling to the scalar and/or the pseudoscalar operators. If that scenario is verified in Nature, it would strongly affect $B\to K \ell^+\ell^-$, whereas the three transverse asymmetries $A_T^{(2,{\rm im},{\rm re})}(q^2)$  of $B\to K^\ast \ell^+\ell^-$ decay would remain unchanged with respect to their shapes predicted at low $q^2$'s in the SM~\cite{damir-elia}.~\footnote{Please note that the asymmetry $A_T^{(2)}(q^2)$ has been introduced in ref.~\cite{Melikhov} and since then abundantly discussed in the literature. } If, instead,  the NP alters the couplings to the semileptonic operators $\propto \bar s\gamma_\mu P_{L,R} b$, then the shapes of the mentioned asymmetries would change too.

In this paper we assume that the NP does not couple with tensor operators (specified below). It turns out that this assumption can too be tested by measuring $A_{FB}^\ell(q^2)$, the forward-backward asymmetry in $B\to K\ell^+\ell^-$,  but in the region of large $q^2$'s in which the non-zero tensor couplings would entail a large enhancement of $A_{FB}^\ell(q^2)$. 

In what follows we will show how and why the two decay modes, $B_s\to\mu^+\mu^-$ and $B\to K \ell^+\ell^-$, are complementary, and after combining the recent experimental results with our current theoretical knowledge of the hadronic matrix elements we will discuss the resulting constraints on NP.

In handling the $B\to K \ell^+\ell^-$ decay we computed the form factors by using the simulations of quenched QCD on the lattice (LQCD), which appear to be compatible with the results obtained by evaluating the QCD sum rules near the light cone (LCSR)~\cite{ball-zwicky}. These results are merely an illustration of the potential of LQCD in pinning down the hadronic errors in this decay. Very soon these results will be substantially improved by using the available QCD configurations that contain the effects of dynamical quarks~\cite{bkk-latt}. Notice that the lattice results are more reliable in  the larger half of the $q^2$-region available from this decay ($14\ \gev^2\lesssim q^2\lesssim  20\ \gev^2$). For that reason it would be desirable to have the experimentally established partial decay width of  $B\to K \ell^+\ell^-$ at $q^2$'s that are also accessible in modern lattice QCD studies. In that way we would be far more confident about the quantitative statements made from this kind of analyses.

The effective Hamiltonian describing the $b\to s\ell^+\ell^-$ transitions at low energy is~\cite{Heff}
 \begin{equation} \label{eq:Heff}
  {\cal H}_{\eff} = - \frac{4\,G_F}{\sqrt{2}} V_{tb}V_{ts}^\ast
     \left[  \sum_{i=1}^{6} C_i (\mu)
\mathcal O_i(\mu) + \sum_{i=7,8,9,10,P,S,T,T5} \biggl(C_i (\mu) \mathcal O_i + C'_i (\mu) \mathcal
O'_i\biggr)\right] \,,
\end{equation}
where the twice Cabibbo suppressed contributions ($\propto  V_{ub}V_{us}^\ast $) have been neglected. 
The operator basis in which the Wilson coefficients have been computed is~\cite{Bobeth:1999mk,Altmannshofer:2008dz}:
\begin{align}\label{basisOps}
{\mathcal{O}}_{7} &= \frac{e}{g^2} m_b
(\bar{s} \sigma_{\mu \nu} P_R b) F^{\mu \nu} ,&
{\mathcal{O}}_{7}^\prime &= \frac{e}{g^2} m_b
(\bar{s} \sigma_{\mu \nu} P_L b) F^{\mu \nu} ,\nn \\
{\mathcal{O}}_{8} &= \frac{1}{g} m_b
(\bar{s} \sigma_{\mu \nu} T^a P_R b) G^{\mu \nu \, a} ,&
{\mathcal{O}}_{8}^\prime &= \frac{1}{g} m_b
(\bar{s} \sigma_{\mu \nu} T^a P_L b) G^{\mu \nu \, a} ,\nn \\
{\mathcal{O}}_{9} &= \frac{e^2}{g^2} 
(\bar{s} \gamma_{\mu} P_L b)(\bar{\ell} \gamma^\mu \ell) ,&
{\mathcal{O}}_{9}^\prime &= \frac{e^2}{g^2} 
(\bar{s} \gamma_{\mu} P_R b)(\bar{\ell} \gamma^\mu \ell) ,\nn \\
{\mathcal{O}}_{10} &=\frac{e^2}{g^2}
(\bar{s}  \gamma_{\mu} P_L b)(  \bar{\ell} \gamma^\mu \gamma_5 \ell) ,&
{\mathcal{O}}_{10}^\prime &=\frac{e^2}{g^2}
(\bar{s}  \gamma_{\mu} P_R b)(  \bar{\ell} \gamma^\mu \gamma_5 \ell) ,\nn \\
{\mathcal{O}}_{S} &=\frac{e^2}{16\pi^2}
 (\bar{s} P_R b)(  \bar{\ell} \ell) ,&
 {\mathcal{O}}_{S}^\prime &=\frac{e^2}{16\pi^2}
 (\bar{s} P_L b)(  \bar{\ell} \ell) ,\nn \\
{\mathcal{O}}_{P} &=\frac{e^2}{16\pi^2}
 (\bar{s} P_R b)(  \bar{\ell} \gamma_5 \ell) ,&
 {\mathcal{O}}_{P}^\prime &=\frac{e^2}{16\pi^2}
 (\bar{s} P_L b)(  \bar{\ell} \gamma_5 \ell),,\nn \\
{\mathcal{O}}_{T} &=\frac{e^2}{16\pi^2}
 (\bar{s} \sigma_{\mu\nu} b)(  \bar{\ell}\sigma^{\mu\nu} \ell) ,&
 {\mathcal{O}}_{T5} &=\frac{e^2}{16\pi^2}
 (\bar{s} \sigma_{\mu\nu} b)(  \bar{\ell} \sigma^{\mu\nu}\gamma_5 \ell),
\end{align}
where $P_{L,R}=(1\mp \gamma_5)/2$, $\ell = e$ or $\mu$, and the explicit expressions for $\mathcal O_{1-6}$ can be found in ref.~\cite{Bobeth:1999mk}.  Short distance physics effects, encoded in the Wilson coefficients $C_i (\mu)$, have been computed in the Standard Model (SM) through a perturbative matching of the effective with the full theory at $\mu=m_W$,  and then evolved down to $\mu=m_b$ by means of the QCD renormalization group equations at next-to-next-to-leading logarithmic approximation (NNLO)~\cite{Bobeth:1999mk}. 
It is customary to reassemble Wilson coefficients multiplying the same hadronic matrix element into effective coefficients appearing in the physical amplitudes, namely~\cite{Wilsoneff},
\bea\label{eq:redef}
C_7^{\rm eff} & = & \frac{4\pi}{\alpha_s}\, C_7 -\frac{1}{3}\, C_3 -
\frac{4}{9}\, C_4 - \frac{20}{3}\, C_5\, -\frac{80}{9}\,C_6\,,
\nonumber\\
C_8^{\rm eff} & = & \frac{4\pi}{\alpha_s}\, C_8 + C_3 -
\frac{1}{6}\, C_4 + 20 C_5\, -\frac{10}{3}\,C_6\,,
\nonumber\\
C_9^{\rm eff} & = & \frac{4\pi}{\alpha_s}\,C_9 + Y(q^2)\,,
\nonumber\\
C_{10}^{\rm eff} & = & \frac{4\pi}{\alpha_s}\,C_{10}\,,\qquad
C_{7,8,9,10}^{\prime,\rm eff} = \frac{4\pi}{\alpha_s}\,C'_{7,8,9,10}\,,
\eea
where the function $Y(q^2)$ is
\bea\label{eq:Y}
Y(q^2) & = &  \frac{4}{3}\, C_3 + \frac{64}{9}\, C_5 + \frac{64}{27} C_6 -\frac{1}{2}\,h(q^2,0) \left( C_3 + \frac{4}{3}\,C_4 + 16 C_5
  + \frac{64}{3}\, C_6\right)\nonumber\\
& &
  \hspace*{-23mm}+ h(q^2,m_c) \left( \frac{4}{3}\, C_1 + C_2 + 6 C_3 + 60 C_5\right)
 -\frac{1}{2}\,h(q^2,m_b) \left( 7 C_3 + \frac{4}{3}\,C_4 + 76 C_5
  + \frac{64}{3}\, C_6\right),
\eea
and
\begin{equation}
h(q^2,m_q) = -\frac{4}{9}\, \left( \ln\,\frac{m_q^2}{\mu^2} - \frac{2}{3}
- z \right) - \frac{4}{9}\, (2+z) \sqrt{|z-1|} \times 
\left\{
\begin{array}{l@{\quad}l}
\displaystyle\arctan\, \frac{1}{\sqrt{z-1}} & z>1\\[10pt]
\displaystyle\ln\,\frac{1+\sqrt{1-z}}{\sqrt{z}} - \frac{i\pi}{2} & z \leq 1
\end{array}
\right. ,
\end{equation}
with $z=4 m_q^2/q^2$.
To make the notation less heavy, in what follows we will drop the superscript ``eff" in Wilson coefficients, while tacitly assuming the redefinitions~(\ref{eq:redef}). Note that in the SM the right-handed flavour violating couplings or to the operators ${\cal O}_{S,P,T,T5}$ are absent and therefore the Wilson coefficients $C_{7,8,9,10}^\prime = C_{S,P,T,T5}^{(\prime)}=0$.  

\section{Physical processes as constraints\label{sec:3}}

\subsection{$B_s \to \ell^+ \ell^- $}
One of the most promising decay modes expected to reveal the effects of NP at LHCb is $B_{s} \to \mu^+ \mu^-$. Using the effective theory~(\ref{eq:Heff}) the only operator contributing to the amplitude of this process in the SM is ${\cal O}_{10}$, and the expression for the 
branching fraction of this decay reads:
\begin{equation}
\begin{split}\label{Bsmm-sm}
{\rm Br} \left( B_s \to \ell^+ \ell^- \right)^{\rm SM} = \tau_{B_s}  { G_F^2 \alpha^2 \over 16 \pi^3 } |V_{tb} V^{\ast}_{ts} |^2  &  m_{B_s}  m_\ell^2 \beta_\ell(m_{B_s}^2)  | C_{10} |^2   f_{B_s}^2 \, ,
\end{split}
\end{equation}
where $\beta_\ell(q^2)=\sqrt{  1- { 4 m_\ell^2/q^2}   }$,  $\ell =\mu$ or $e$,  and the $B_s$ meson decay constant is defined via
\bea
\langle 0 \vert  \bar{s}  \gamma_{\mu} P_L b \vert B_s(p)\rangle ={i\over 2} f_{B_s} p_\mu\,.
\eea
On general grounds the contributions of physics beyond the Standard Model (BSM) can modify the above expression to 
\begin{equation}\label{Bsmm-bsm}
\begin{split}
{\rm Br} \left( B_s \to \ell^+ \ell^- \right)^{\rm BSM} = \tau_{B_s} f_{B_s}^2 m_{B_s}^3 { G_F^2 \alpha^2 \over 64 \pi^3  } & |V_{tb} V^{\ast}_{ts} |^2   \beta_\ell(m_{B_s}^2)  \left[ \
{m_{B_s}^2 \over m_b^2}\Big| C_S -C_S^\prime \Big|^2  \left(1- { 4 m_\ell^2 \over m_{B_s}^2}  \right) \right.  \\ 
 & \,   \left. +  \ \Big| {m_{B_s}  \over m_b } \left( C_P -C_P^\prime \right) + 2 {   m_\ell \over m_{B_s}} \left( C_{10} - C_{10}^{\prime  } \right) \Big|^2 \, \right] \, ,
\end{split}
\end{equation}
thus lifting the helicity suppression exhibited in the SM expression~(\ref{Bsmm-sm}). This is why it is interesting and important to experimentally investigate this decay both for $\ell =\mu$ and $\ell=e$. In the SM the electron mass severely suppresses the decay rate, and a clean detection of the $B_s\to e^+e^-$ events would be a clear signal of NP effects. In the last formula the presence of the right handed (non-SM) couplings would induce $C_{10}^\prime\neq 0$, while the new scalar (pseudoscalar) couplings would entail the non-zero  $C_S^{(\prime)}$ ($C_P^{(\prime)}$). 
\begin{figure}[t!]
\begin{center}
{\resizebox{11cm}{!}{\includegraphics{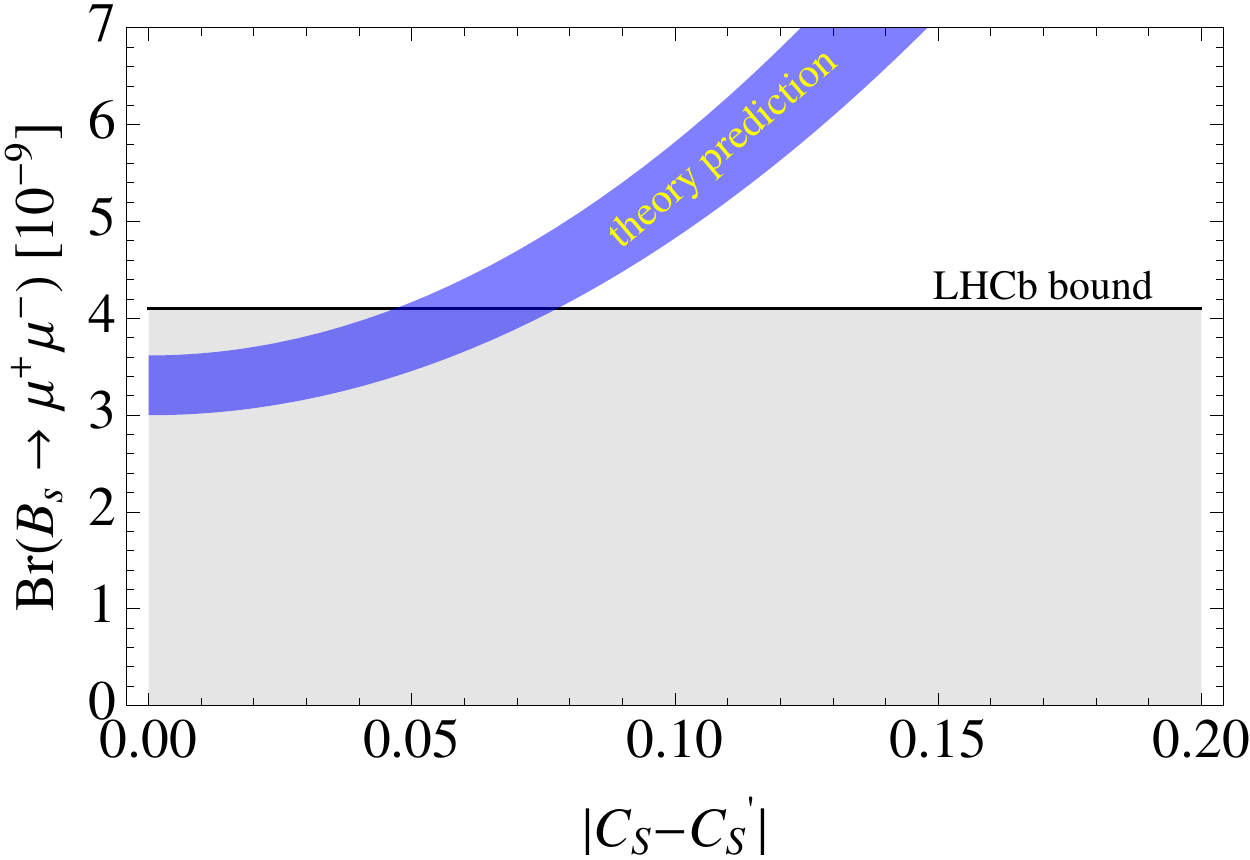}}} 
\caption{\label{fig:1}\footnotesize{\sl 
Increase of the ${\rm Br} \left(B_{s} \to \mu^+ \mu^- \right)$ with  $ \left| C_{S} - C_{S}^{\prime} \right|$ according to eq.~(\ref{Bsmm-bsm}) is shown by the dark band defined by $1\sigma$ errors on the input parameters. The brightly shaded area depicts the experimentally allowed values for the decay mode. $C_{10}^{\prime}= C_P^{(\prime)}=0$ has been used.
}} 
\end{center}
\end{figure}

Several specific NP models that allow for large values of $C_S$
suggested a possibility of observing a large number of $B_{s} \to
\mu^+ \mu^- $ events. Recent experimental activity at the
LHC~\cite{LHCb-Bmumu}, however, showed that such a large enhancement
does not occur and the current upper bound on ${\rm Br} \left( B_s \to
  \mu^+ \mu^- \right)$ is quite close to the SM value. More
specifically,\footnote{Recently, it has been noted that the effect of
  $B_s-\bar B_s$ mixing should be taken into account~\cite{Fleischer} and whose net
  effect amounts to replacing $\mathrm{Br}(B_s \to \mu^+ \mu^-)^{\rm exp} 
  \to  (1-y_s) \mathrm{Br}(B_s \to \mu^+ \mu^-)^{\rm exp}$, where $y_s
  = (\Delta\Gamma/2\Gamma)_{B_s}
  = (9.0 \pm 2.1 \pm 0.8)\,\%$ was also measured by LHCb~\cite{1202.4717}. This
  correction is already incorporated in the corrected experimental value~\eqref{bsmumu-exp}.} 
\bea\label{bsmumu-exp}
&&{\rm Br} \left(B_{s} \to \mu^+ \mu^- \right)^{\rm exp} <    4.1 \times 10^{-9} \,, \cr
&&\\ 
&& {\rm Br} \left(B_{s} \to \mu^+ \mu^- \right)^{\rm th-SM} =  (3.3 \pm 0.3 )\times 10^{-9} \nn\, ,
\eea
where for the SM estimate we used the parameters given in the Appendix~B of the present paper (c.f. tab.~\ref{tab:1}). Although the desired enhancement by orders of magnitude with respect to the SM  does not occur, the possibility of seeing the NP signal from this decay mode is still alive. One can even envisage the possibility of  ${\rm Br} \left(B_{s} \to \mu^+ \mu^- \right)$ smaller than the one predicted in the SM, as it can be easily seen from eq.~(\ref{Bsmm-bsm}).

If, for the moment, we only consider the possibility that the SM is extended by allowing a coupling to the scalar operator, i.e. by keeping  $C_{10}^\prime= C_P^{(\prime)}= 0$, then from eq.~(\ref{Bsmm-bsm}) one gets 
\bea\label{constr-cSmoins}
 \left| C_{S} - C_{S}^{\prime} \right| \leq 0.08 \qquad (1\ \sigma)\,,
\eea
that we illustrate in fig.~\ref{fig:1}. This limit is actually not too far away from $\left| C_{S} - C_{S}^{\prime} \right|^{\rm SM} = 0$, which is why this bound becomes a very severe constraint on the NP models with scalar operators.  The above bound still allows for an observation of  ${\rm Br}(B_s \to e^+e^-)\lesssim 2\times 10^{-9}$, unless the lepton flavor universality is not respected by NP.  Obviously, the bound~(\ref{constr-cSmoins}) does not give us any information about $C_{S}$ and $C_{S}^\prime$ separately. For that we would need a complementary information that can be inferred from $B\to K\ell^+\ell^-$. 

Recent research about exploring the constraint from the experimental bound on ${\rm Br}(B_s\to \mu^+\mu^-)$ has been reported in ref.~\cite{RECENT}.

\subsection{\label{sectionKll}$B \to K\ell^+ \ell^- $}

There are many papers in the literature dealing with this decay. We were able to check the expressions given in ref.~\cite{Bobeth:2007dw} with which we agree. 
The full distribution of this decay is conveniently expressed as,
\bea\label{WidthBK}
 \frac{ d^2 \Gamma_\ell (q^2, \cos \theta) }{dq^2 d \cos \theta } =  a_\ell \qq + b_\ell \qq \cos \theta + c_\ell \qq \cos^2 \theta   \,,
\eea
where for short $\Gamma_\ell \equiv \Gamma(B \to K\ell^+ \ell^-)$, $q^2=(p_{\ell^+}+p_{\ell^-})^2$, and $\theta$ is the angle between the directions of $\bar B$ and of $\ell^-$ in the center of mass frame of the lepton pair. In terms of specific Lorentz components, $F_{i}(q^2)$ ($i=S,P,V,A,T,T5$), the explicit expressions of the functions on the right hand side of eq.~(\ref{WidthBK}) are:
\begin{align}\label{coeff_BKell}
a_\ell \qq = &\  {\cal C}(q^2)\Big[  q^2 \left( \beta_{\ell}^2(q^2) \lvert F_S(q^2)\rvert^2 + \lvert F_P(q^2) \rvert^2 \right)  +
 \frac{\lambda \qq }{4}  \left( \lvert F_A (q^2)\rvert^2 +  \lvert F_V(q^2) \rvert^2 \right)   \nn \\ 
 & \hspace{2.25cm}+ 4 m_{\ell}^2 m_B^2 \lvert F_A(q^2) \rvert^2+2m_{\ell}  \left( m_B^2 -m_K^2 +q^2\right) \text{Re}\left( F_P(q^2) F_A^{\ast}(q^2) \right)\Big]  \,,\nn \\ 
&\nn\\
b_\ell \qq = &\  2\  {\cal C}(q^2)\Big\{q^2 \left[  \beta_{\ell}^2(q^2)  \text{Re}  \left( F_S(q^2) F_T^{\ast}(q^2) \right) + \text{Re} \left(F_P(q^2) F_{T5}^{\ast}(q^2)  \right)\right] \nn \\
 & +    m_{\ell}  \big[\sqrt{\lambda \qq } \beta_{\ell}(q^2) \text{Re} \left( F_S(q^2) F_V^{\ast}(q^2)\right) +\left(m_B^2 - m_K^2+q^2 \right) \text{Re} \left(F_{T5}(q^2) F_A^{\ast}(q^2) \right) \big] \Big\} \,, \nn \\
&\nn\\
c_\ell \qq = &\  {\cal C}(q^2)\Big[ q^2 \left( \beta_{\ell}^2(q^2) \lvert F_T(q^2) \rvert^2 + \lvert F_{T5}(q^2) \rvert^2 \right)  - \frac{\lambda \qq}{4} \beta_{\ell}^2(q^2) \left( \lvert F_A(q^2) \rvert^2 +  \lvert F_V(q^2) \rvert^2 \right)  \nn \\ 
 & \hspace{4.25cm}+ 2 m_{\ell} \sqrt{\lambda \qq} \beta_{\ell}(q^2) \text{Re} \left( F_T(q^2) F_V^{\ast}(q^2) \right) \Big]\,,
\end{align}
where 
\bea
{\cal C}(q^2)= \frac{G_F^2 \alpha^2 \lvert V_{tb} V_{ts}^{\ast} \rvert^2}{512 \pi^5 m_B^3}  \beta_{\ell}(q^2)   \sqrt{\lambda \qq } \,,
\eea
with 
\bea
\lambda \qq = q^4+ m_B^4 + m_K^4 -2 \left( m_B^2 m_K^2 +m_B^2 q^2 + m_K^2 q^2 \right) \,,
\eea
and 
\bea\label{F_BKell}
F_V \qq  &=&   \left( C_9 + C_9^{\prime  } \right)  f_+ \qq + \frac{2 m_b}{m_B +m_K} \left(   C_7  + C_7^{\prime }  + {4 m_{\ell}\over m_b}  C_T \right)  f_T \qq   \,,  \nn \\ 
F_A \qq    &=&  \left( C_{10} +C_{10}^{\prime  } \right)  f_+ \qq \,,   \nn\\
 F_S \qq &=&     \frac{m_B^2 -m_K^2}{2m_b}  \left( C_S + C_S^{\prime} \right) f_0 \qq\,,  \nn \\
F_P \qq  &=&  \frac{m_B^2 -m_K^2}{2m_b} \left( C_P + C_P^{\prime} \right)  f_0 \qq  \nn \\
&& - m_{\ell} \left( C_{10}+C_{10}^{\prime } \right) \left[  f_+ \qq - \frac{m_B^2-m_K^2}{q^2} \left(f_0  \qq - f_+ \qq \right) \right]  \,,\nn\\
F_T \qq &=&   \frac{2 \sqrt{\lambda \qq} \beta_{\ell}(q^2)}{m_B + m_K}  C_T f_T\qq\,,\nn\\
 F_{T5} \qq  &=&  \frac{2 \sqrt{\lambda \qq} \beta_{\ell}(q^2)}{m_B + m_K}  C_{T5}f_T\qq \,. 
\eea
In the above expressions we employed the standard decompositions of the hadronic matrix elements in terms of the form factors, namely,
\bea\label{matrix-elements}
\langle K(k)| \bar{s} \gamma_{\mu} b |B(p)\rangle & = & \left[ (p + k)_\mu - {m_B^2 - m_K^2 \over q^2} q_\mu \right] f_{+}\qq + {m_B^2-m_K^2 \over q^2} q_{\mu} f_{0}\qq \, , \nn \\
\langle K(k)| \bar{s}\sigma_{\mu\nu}b |B(p) \rangle &=& -i \left( p_\mu k_\nu  - p_\nu k_\mu \right) \frac{2 f_T\qq}{m_B + m_K} \, .
\eea
After integrating eq.~(\ref{WidthBK}) over $q^2=(p-k)^2$, one obtains 
\bea\label{angDistr}
{ d \Gamma_\ell ( \cos \theta) \over d \cos \theta } = A_\ell + B_\ell \cos \theta + C_\ell \cos^2 \theta \,,
\eea
where obviously
\begin{align}
 A_{\ell} & = \int_{q_{\rm min}^2}^{q_{\rm max}^2} a_\ell \qq d q^2\, , & B_{\ell} & = \int_{q_{\rm min}^2}^{q_{\rm max}^2} b_\ell  \qq d q^2\, , & C_{\ell} & = \int_{q_{\rm min}^2}^{q_{\rm max}^2} c_\ell  \qq d q^2\, ,
\end{align}
with $q_{\rm min}^2=4m_\ell^2$,  and $q_{\rm max}^2=(m_B-m_K)^2$. For a partial decay width, one would obviously choose different  $q_{\rm min/max}^2$. Finally,  the integration over $\theta$ leads to the full decay width, in which the term proportional to $\cos\theta$ drops out. The  latter survives in the forward-backward asymmetry and we have,
\bea \label{observables-Kll}
 \Gamma_{\ell} & = 2 \left( A_{\ell} + \frac{1}{3} C_{\ell} \right) \, ,    \qquad  A_{FB}^{\ell} & =\frac{B_\ell}{\Gamma_\ell} \, .
\eea
Clearly these two observables are independent as they involve different pieces of the distribution~(\ref{WidthBK}).  In ref.~\cite{Bobeth:2007dw} another useful quantity has been introduced,  
\bea
F^\ell_{H} = {2\left(A_\ell + C_\ell \right)\over \Gamma_\ell}\,,
\eea
 which in the SM is proportional to $m_\ell^2$, but can receive important contributions in various scenarios of NP. Before continuing, we wish to stress that in the SM $A_{FB}^{\ell}=0$, and it remains zero even if the Wilson coefficients $C_{7,9,10}^{(\prime)}$ received large NP contributions.  Its non-zero measurement would be a clean signal of NP and therefore its experimental study would be highly welcome.

\begin{figure}[t!]
\begin{center}
{\resizebox{11cm}{!}{\includegraphics{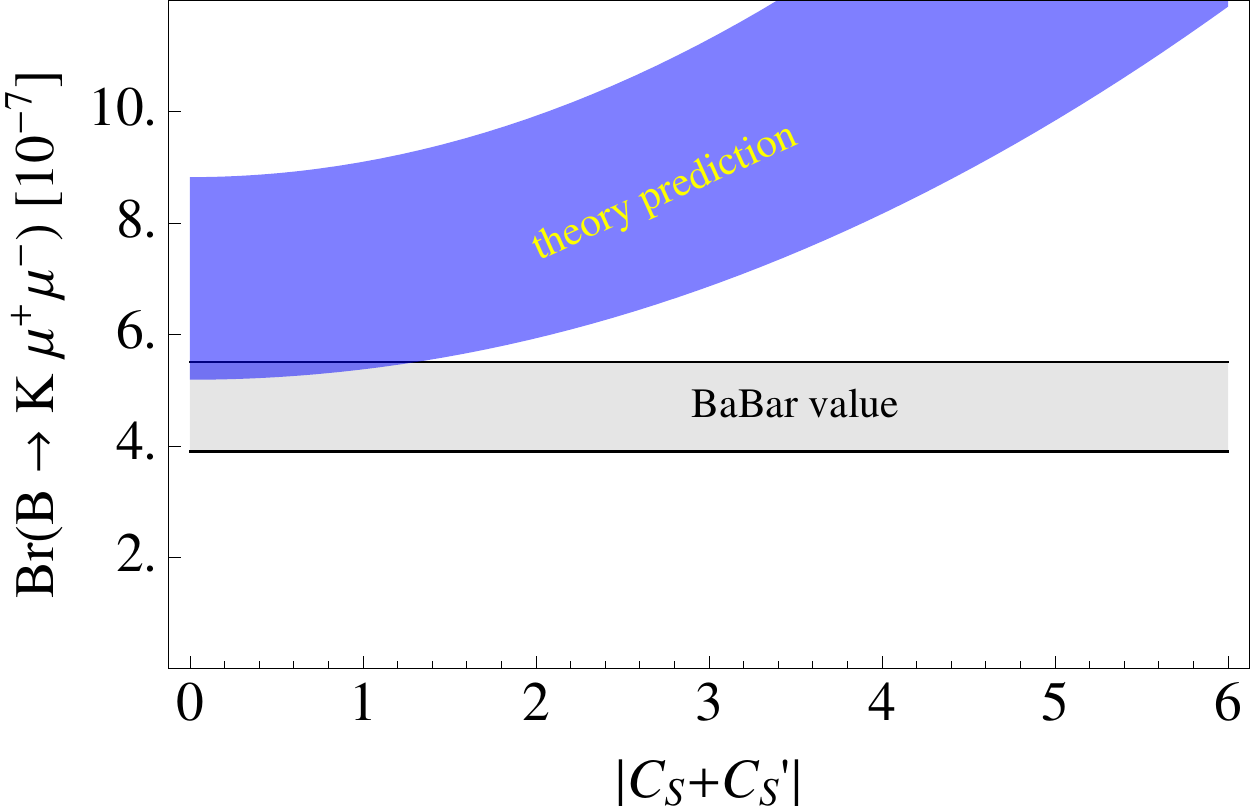}}} 
\caption{\label{fig:2}\footnotesize{\sl ${\rm Br} \left(B \to K\ell^+
      \ell^- \right) = \tau_B \Gamma_\ell$ as a function of $ \left|
      C_{S} + C_{S}^{\prime} \right|$. Bright shaded horizontal band
    corresponds to the recently measured ${\rm Br} \left(B \to K\ell^+
      \ell^- \right)$ at BaBar. We used $ \Gamma_\ell$ given in
    eq.~(\ref{observables-Kll}), and $\tau_B= \tau_{B^0}$, to conform
    with the experimental practice when combining neutral and
    charged $B$-modes.}}
\end{center}
\end{figure}

For our purpose, it is important to note  that all the functions in eqs.~(\ref{coeff_BKell},\ref{F_BKell}) involve the sum of $C_i+C_i^\prime$, and therefore the full branching ratio provides us with a constraint that is complementary to eq.~(\ref{constr-cSmoins}). The recently measured~\cite{exp-BKll}
\bea\label{Kll-3}
&&{\rm Br} \left(B \to K\ell^+ \ell^- \right)_{\rm exp} =   (4.7 \pm 0.6\pm 0.2) \times 10^{-7}  \,,
\eea
is compatible with the SM prediction,
\bea\label{Kll-sm} {\rm Br} \left(B \to K\ell^+ \ell^- \right)_{\rm SM}  = \left\{ \begin{array}{ll}
         (7.5 \pm 1.4 )\times 10^{-7} & \mbox{\rm LQCD}\,,\\
         &\\
         (6.8 \pm 1.6 )\times 10^{-7} & \mbox{\rm LCSR}\,.\end{array} \right., \eea 
 which, at this stage, we take to be
 \bea
 {\rm Br} \left(B \to K\ell^+ \ell^- \right)_{\rm SM}  = (7.0 \pm 1.8 )\times 10^{-7}\,,
 \eea
thus covering all the values allowed by the two methods of computing the form factors, LQCD and LCSR.
Allowing for non-zero $C_{S}^{(\prime)}$ then leads to 
\bea\label{constr-cSplus}
 \left| C_{S} + C_{S}^{\prime} \right| \leq 1.3 \qquad (1\ \sigma)\,,
\eea
as illustrated in fig.~\ref{fig:2}. To obtain eq.~(\ref{constr-cSplus})  we had to assume that NP does not alter the SM values of $C_{7,9}^{(\prime)}$ which enter the expression for ${\rm Br} \left(B\to K \ell^+\ell^-\right)$. That assumption can be tested experimentally through the study of low-$q^2$ behavior of three transverse asymmetries discussed in ref.~\cite{damir-elia} which exhibit three very important features: (i) they have small hadronic uncertainties, (ii) their shapes are highly sensitive to $C_{7,9}^{(\prime)}$, and (iii) they are completely insensitive to $C_{S,P}^{(\prime)}$. 

Another important comment concerning the constraint~(\ref{constr-cSplus}) is that it is obtained by including the $1 \sigma$ experimental uncertainty,  and by using the form factors that are either obtained from the numerical simulations of quenched QCD on the lattice (see Appendix~B of this paper), or in the QCD sum rule analysed near the light cone~\cite{ball-zwicky}, respectively labelled as LQCD and LCSR in eq.~(\ref{Kll-sm}). LQCD results appear to be consistent with those obtained from LCSR. The three relevant form factors, $f_{+,0,T}(q^2)$, will soon be improved by the new generation of unquenched lattice QCD simulations. Several such studies are underway~\cite{bkk-latt}. Notice that the improvement of $f_{+,0,T}(q^2)$ is much more realistic to expect than those parameterizing the $B\to K^\ast$ transition matrix elements, because the latter decay involves many more form factors, including at least three that suffer from very large uncertainties (see e.g. ref.~\cite{our-vec}).

\section{Constraints on the Scalar (Pseudoscalar) Couplings}

Couplings to the scalar and pseudoscalar operators are particularly interesting in the framework of supersymmetric (SUSY) extensions of the Standard Model.  A very detailed consideration of the SUSY contributions to $B_s\to \ell^+\ell^-$ has been made in ref.~\cite{Chankowski:2000ng}. The relation between the Wilson coefficients $C_{S,P}^{(\prime)}$ and $C^S_{LL,LR,RL,RR}$, defined in ref.~\cite{Chankowski:2000ng}, are:
\bea\label{eq:chank}
C_S = X \left(C_{LR}^S + C_{LL}^S\right)^\ast\,,&&
C_P = X \left(C_{LR}^S - C_{LL}^S\right)^\ast\,,\nn\\ 
C_S^\prime = X \left(C_{RR}^S + C_{RL}^S\right)^\ast\,,&&
C_P^\prime = X \left(C_{RR}^S - C_{RL}^S\right)^\ast\,,
\eea
where $X=\pi/(\sqrt{2}  G_F \alpha V_{ts}^\ast V_{tb})$. From ref.~\cite{Chankowski:2000ng} we learn that the SUSY contributions to the box and penguin diagrams can modify $C_{S,P}^{(\prime)}$ as follows:
\begin{itemize}
\item Diagrams with one charged Higgs boson propagating in the box can give non-zero contribution via coupling to the left handed parts only. Furthermore, they verify $C_S^{H^+}= -C_P^{H^+}$.  The right-handed couplings, instead, are suppressed by the strange quark mass,  $C_S^{\prime\  H^+} = C_P^{\prime\  H^+} =0$. 
\item Diagrams with charginos propagating in the box also give rise to $C_S^\chi$ and $C_P^\chi$, but leave $C_S^{\prime\ \chi} = C_P^{\prime\ \chi} =0$. Moreover, if for example the masses of squark and sneutrino in the box are degenerate, then one again obtains $C_S^{\chi} =- C_P^{\chi}$. 
\item $Z^0$-penguin diagram, with super-particles propagating in the loop, cannot generate a contribution to $C_{S,P}^{(\prime)}$ due to the vector coupling to $Z^0$.
\item $H^0$-penguin, instead, can give a sizable contribution which verifies 
\bea
C_S^{H^0} = -C_P^{H^0}\,, \quad C_S^{\prime H^0} =  C_P^{\prime  H^0 }\,,
\eea
up to the electro-weak symmetry breaking corrections that are proportional to the mass splitting between the SUSY (MSSM) Higgs bosons, namely $m_{H^0}^2-m_{A^0}^2$.  
\end{itemize}
The situation similar to the $H^0$-penguin case above also happens in the models with vector leptoquark states. In those models the non-zero contributions to $C_S$ are possible and they satisfy the relation $C_S=\pm C_P$, as well as $C_S^\prime = \mp C_P^\prime$~\cite{nejc}.  On the contrary, the models with  scalar leptoquarks cannot generate any sizeable value of $C_{S,P}^{(\prime)}$.  These are obviously only two among many scenarios of physics BSM in which the coupling to scalar (pseudoscalar) can be generated.

In the remainder of this section we will combine the constraints discussed in Sec.~\ref{sec:3} and apply them to two particular scenarios: (1) $C_{S}^{(\prime)}\neq 0$, while  $C_{10}^{\prime}=C_{P}^{(\prime)}= 0$, and (2) $C_{P}^{(\prime)}\neq 0$, while  keeping $C_{10}^{\prime}=C_{S}^{(\prime)}= 0$. In each case we will also allow the non-zero relative phase between the left- and right-handed couplings (Wilson coefficients).

\subsection{$C_{S}^{(\prime)}\neq 0$}

\begin{figure}[b!]
\begin{center}
{\resizebox{5.2cm}{!}{\includegraphics{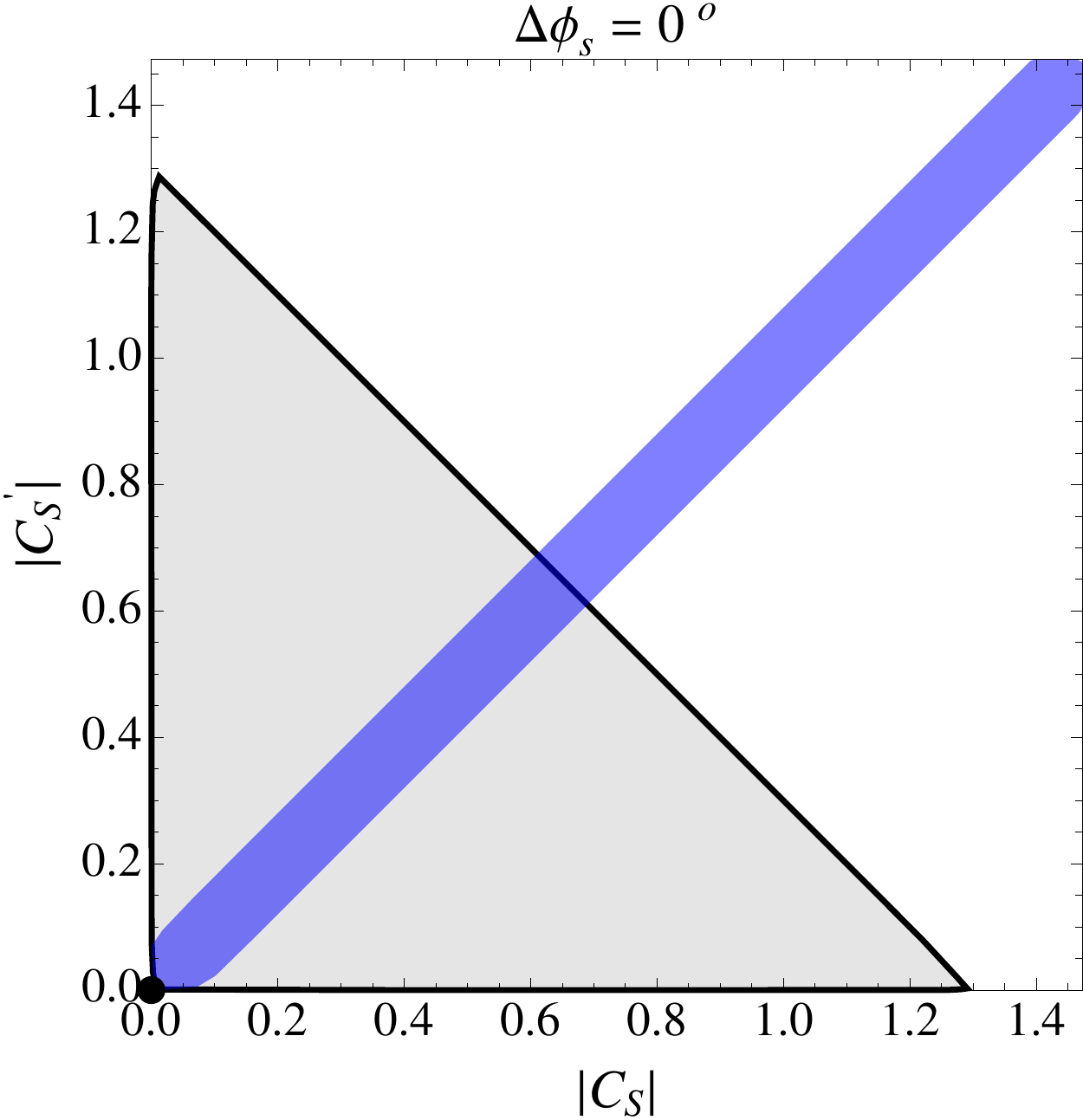}}} 
{\resizebox{5.2cm}{!}{\includegraphics{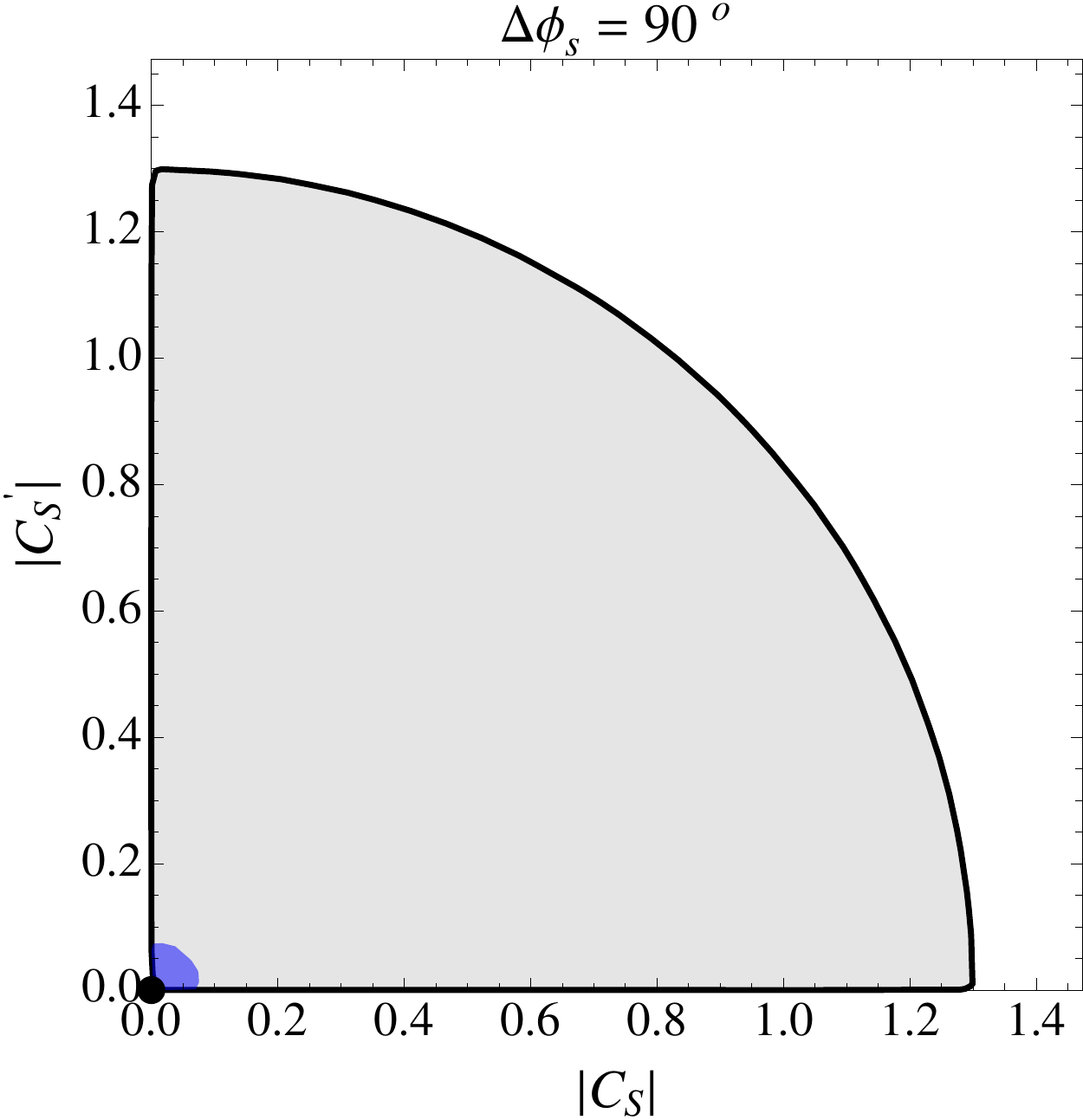}}} 
{\resizebox{5.2cm}{!}{\includegraphics{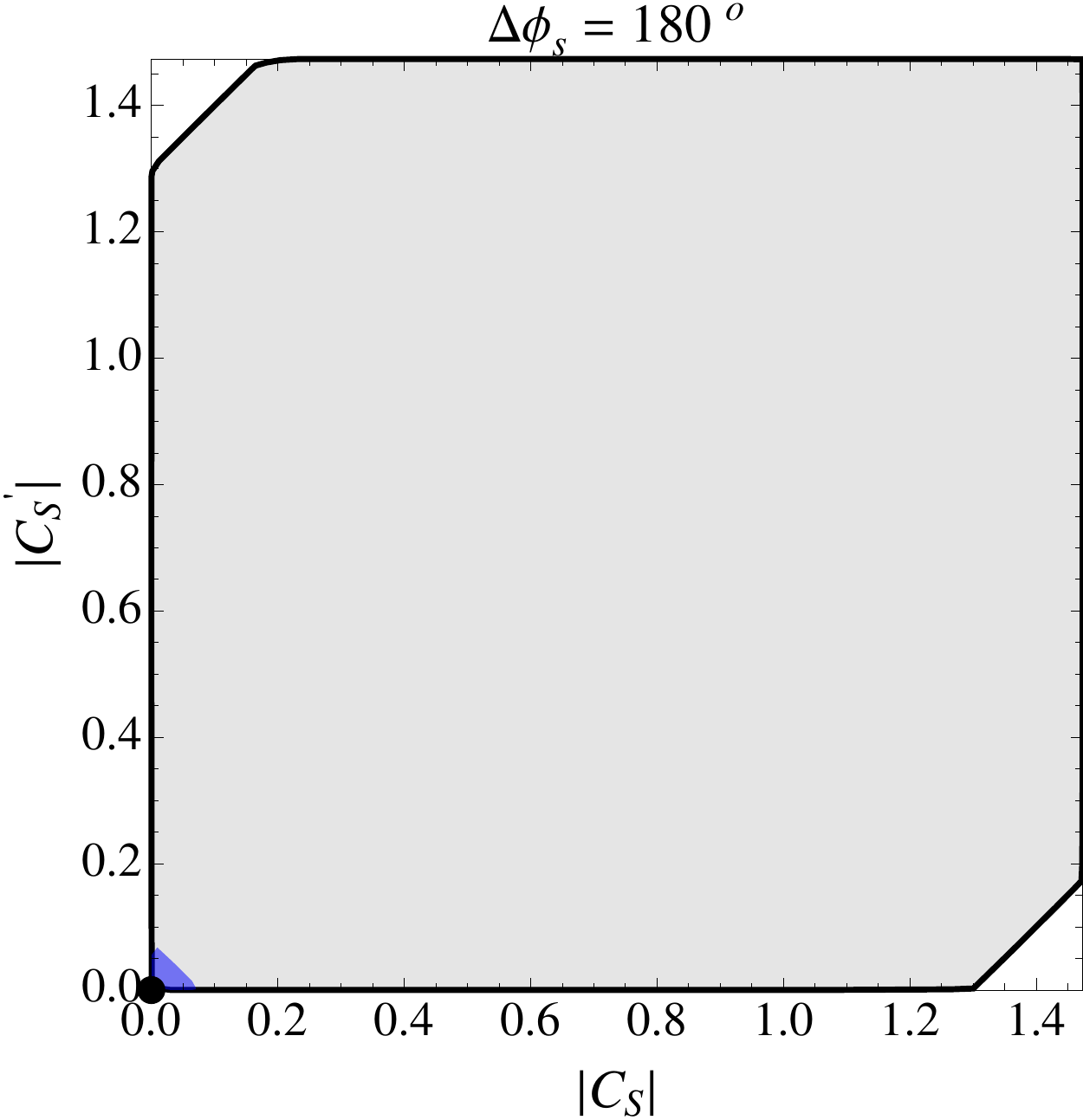}}} 
\caption{\label{fig:5}\footnotesize{\sl 
Constraint on $ \left| C_{S} \right|$  and $ \left|  C_{S}^{\prime} \right|$ obtained from ${\rm Br} \left(B \to K\ell^+ \ell^- \right)$ are represented by the brightly shaded area, whereas the one 
deduced from ${\rm Br} \left(B_s \to \mu^+ \mu^- \right)$ is described by the dark shaded region. Of course, only the overlap of both regions is consistent with the constraints. It is also consistent with ${\rm Br} \left(B \to X_s\ell^+ \ell^- \right)$ and $|C_S^{(\prime)}|\neq 0$ it does not modify the transverse asymmetries in $B\to K^\ast \ell^+\ell^-$. Plotted are the cases with three specific choices of the relative phase $\Delta \phi_S$. }} 
\end{center}
\end{figure}

\begin{figure}[b!]
\begin{center}
{\resizebox{8cm}{!}{\includegraphics{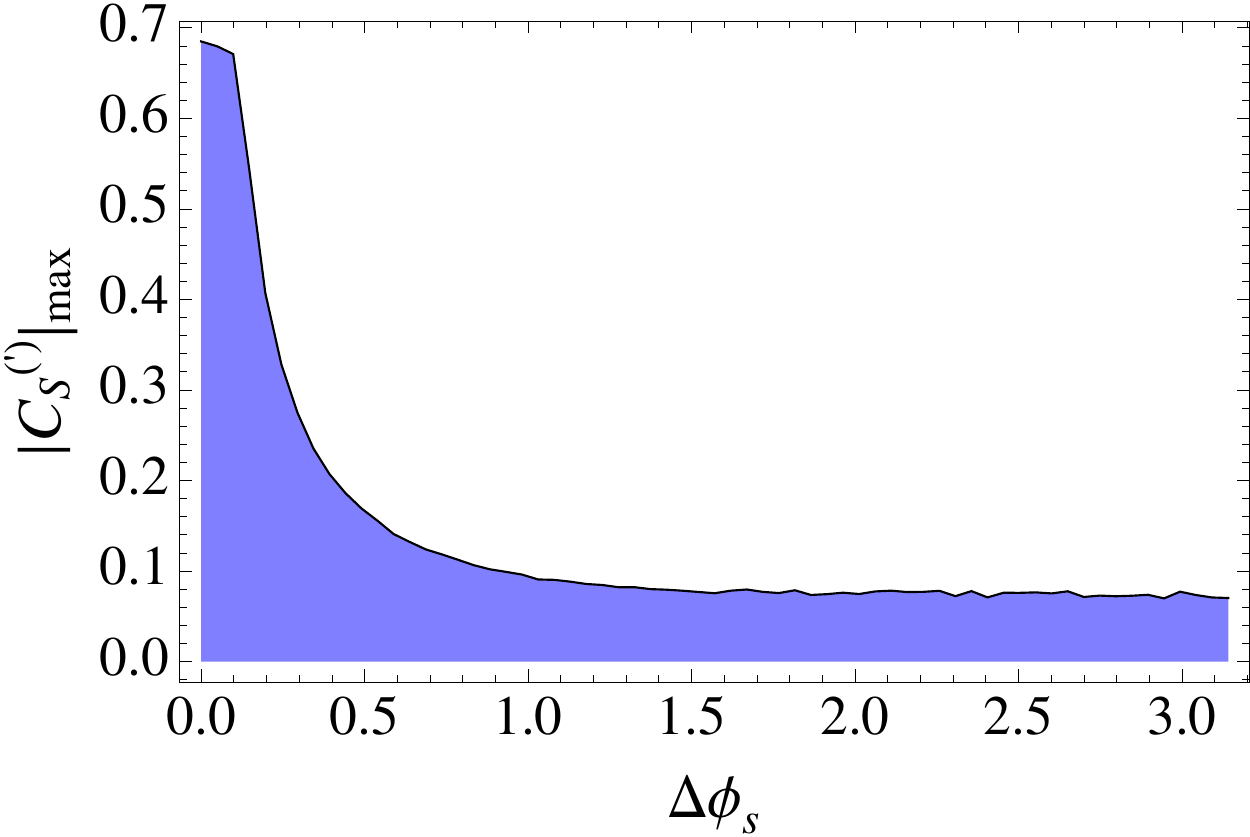}}} 
\caption{\label{fig:6}\footnotesize{\sl 
Maximal value of $ \left| C_{S} \right|$  or $ \left|  C_{S}^{\prime} \right|$ allowed by the constraints~(\ref{constr-cSmoins}) and ~(\ref{constr-cSplus}) as a function of the relative phase $\Delta\phi_S\in [0,\pi]$.}} 
\end{center}
\end{figure}

We first focus on the scenario in which $C_{S}$ and $C_{S}^{\prime}$ can be different from zero. Besides $\vert C_S^{(\prime)}\vert \neq0$, NP can induce new weak phases, which through our constraints in eqs.~(\ref{constr-cSmoins},\ref{constr-cSplus}) cannot be studied separately. Instead, one can study the impact of the relative phase, $\Delta \phi_S=\phi_S^\prime - \phi_S$, by using 
\bea
&&\left| C_{S} \pm C_{S}^{\prime  } \right|^2 =  \left| C_{S}\right|^2 + \left| C_{S}^{\prime  } \right|^2\pm 2  \left| C_{S}\right|  \left| C_{S}^{\prime  } \right| \cos(\Delta \phi_S)\,,
\eea
and explore the possible values of $ \left| C_{S}\right|$ and $ \left| C_{S}^{\prime  }\right|$ for various $\Delta\phi_S \in [0, \pi]$ that are compatible with the constraints (\ref{constr-cSmoins}) and (\ref{constr-cSplus}). The bound on $B_s \to \mu^+ \mu^-$ is much more compelling, and for any non-zero relative phase, $\Delta\phi_S$, the constraint provided by $B \to K\ell^+ \ell^-$ becomes essentially superfluous, except in the case of  $\Delta\phi_S=0$ when the constraint~(\ref{constr-cSmoins}) describes a stripe in the plane ($ \left| C_{S}\right|$, $ \left| C_{S}^{\prime  }\right|$) that is cut off by the constraint~(\ref{constr-cSplus}). This situation is illustrated in fig.~\ref{fig:5} for three representative cases, $\Delta\phi_S = 0, \pi/2$, and $\pi$.  
We see that the situation in which  $\Delta\phi_S=0$ is indeed peculiar, and only for very small values of the relative phase $\Delta\phi_S$ the sizable couplings to new physics via scalar operators are possible. Otherwise a non-zero relative phase entails a reduction of available $\vert C_S^{(\prime)}\vert$, as we show in fig.~\ref{fig:6}.
Two important comments are in order:
\begin{itemize}
\item Any value of $ \left| C_{S} \right|$ and/or $ \left|  C_{S}^{\prime} \right|$ allowed by the constraints (\ref{constr-cSmoins}) and (\ref{constr-cSplus}), for any value of the relative phase $\Delta\phi_S$, are consistent with the 
branching fraction of the inclusive $B\to X_s\ell^+\ell^-$ decay, which has been measured at the $B$-factories at low $q^2$'s~\cite{belle-incl,babar-incl}, resulting in an average~\cite{Huber:2007vv},
\bea\label{bsll-exp}
\int_{1\ \gev^2}^{6\ \gev^2}{d {\rm Br} \left(B \to X_s \mu^+ \mu^- \right) \over dq^2}dq^2 \Big|_{\rm exp}= 1.6(5) \times 10^{-6}\,.
\eea
By using the formulas presented in refs.~\cite{Huber:2007vv,lunghi-hiller}, we obtain in the SM
\bea
\int_{1\ \gev^2}^{6\ \gev^2}{d {\rm Br} \left(B \to X_s \mu^+ \mu^- \right) \over dq^2}dq^2 \Big|_{\rm th-SM}=1.59(17)  \times 10^{-6}\, ,
\eea
where, instead of the usual practice to normalize by the inclusive semileptonic $\Gamma(B\to X_c e\nu)$ decay, we actually use the tree-level decay rate proportional to the fifth power of  the $b$-quark mass, i.e.
\bea
B_0= \tau_B \ {4\ \alpha^2\  G_F^2\ \vert V_{tb}V_{ts}^\ast\vert^2 \  m_b^5\over 3\  (4 \pi)^5  } = 3.41(47)\times 10^{-7}\ ,
\eea 
which is now possible thanks to the fact that the value of the $b$-quark is by now very well determined from the multitude of techniques of QCD sum rules and  modern simulations of QCD on the lattice. Our result agrees very well with the one obtained in ref.~\cite{Huber:2007vv} that was obtained by normalizing to $\Gamma(B\to X_c e\nu)$. In practice we take the average quark mass quoted by PDG and converted it to the pole mass using the NNLO perturbative QCD corrections~\cite{chetyrkin}. 

The scalar contribution to $B \to X_s\mu^+\mu^-$ has been computed in ref.~\cite{Fukae:1998qy}, and leads to a term that adds up to ${\rm Br} \left(B\to X_s\ell^+\ell^- \right)_{\rm SM}$ and reads,  
\bea
 {d {\rm Br} \left(B \to X_s \mu^+ \mu^- \right) \over dq^2}\Big|_{C_S^{(\prime)}} = {3 B_0\over 2  m_b^2 } \left( 1 - {q^2\over  m_b^2}\right)^2  {q^2\over m_b^2}\, \left( | C_S|^2 + |C_S^{\prime}|^2 \right)  \,.
 \eea
When integrated between $1$ and $6\ \gev^2$, this leads to a tiny correction, 
\begin{align}\label{eq:inclS}
\int_{1\ \gev^2}^{6\ \gev^2}{d {\rm Br} \left(B \to X_s \mu^+ \mu^- \right) \over dq^2}dq^2 = 1.59(17)  \times 10^{-6} \left[ 1 + 0.007(1) \left( | C_S|^2 + |C_S^{\prime}|^2 \right)
\right],
\end{align}
so that even the largest allowed values for $\vert C_S^{(\prime)}\vert$, displayed in fig.~\ref{fig:6}, result in a negligibly small correction to the inclusive $B\to X_s\mu^+\mu^-$ rate.
\item We stress again that the NP scenario in which only the coupling to the scalar operators ${\cal O}^{(\prime)}$ are allowed would not modify the SM predictions of the $q^2$-shapes of three transverse asymmetries that can be measured from $B\to K^\ast \ell^+\ell^-$ decay, namely $A_T^{(2,{\rm im},{\rm re})}(q^2)$~\cite{damir-elia}.
\end{itemize}

\subsection{$C_{P}^{(\prime)}\neq 0$}

The situation is slightly more complicated in the case of $C_P \pm C_P^\prime$ because both phases are needed. It is easy to see that eq.~(\ref{Bsmm-bsm}) gives us 
\bea
&&\left| C_P - C_P^\prime + 2 { m_b m_\mu \over m_{B_s}^2}  C_{10}^{\rm SM} \right| \leq \frac{8 \pi^{3/2} m_b}{\alpha  G_F  f_{B_s}
 m_{B_s}^2 |V_{tb} V_{ts}^\ast |}
\sqrt{\frac{\mathrm{Br}(B_s\to \mu^+ \mu^-)}{m_{B_s} \tau_{B_s}
   \beta_\mu(m_{B_s}^2)}} = 0.150\,,
\eea
and therefore considering only the relative phase, $\Delta \phi_P= \phi_{P^\prime}-\phi_P$, is not enough. Instead, we write
\begin{align}\label{eq:xx}
|C_P|^2+|C_P^\prime |^2 &- 2 |C_P| |C_P^\prime | \cos(\Delta \phi_P) \cr
&\hfill \cr
&+ \widetilde C_{10}^2 + 2\ \widetilde C_{10} \bigl[ |C_P| \cos \phi_P -  |C_P^\prime | \cos (\phi_P +\Delta \phi_P)\bigr] \leq (0.150)^2,
\end{align}
where, for brevity, we use $ \widetilde C_{10}= 2  C_{10}^{\rm SM} m_b m_\mu/m_{B_s}^2$. Besides $\Delta \phi_P$ we choose to vary the phase of $C_P$. A compact analytical expression similar to eq.~(\ref{eq:xx}) that  constrains $C_P +  C_P^\prime$ from $\mathrm{Br}(B\to K\ell^+ \ell^-)$ cannot be obtained. Instead, we get,
\bea
&&\left| C_P+ C_P^\prime \right|_{m_\ell=0} \leq  1.3\,,\qquad \left| C_P+ C_P^\prime - 0.33\right|_{m_\mu} \leq  1.3  \quad (1\ \sigma)\,,
\eea
in the massless and massive lepton case respectively. In what follows we will use $m_\ell = m_\mu\neq 0$.

\begin{figure}[h!]
\begin{center}
{\resizebox{5.2cm}{!}{\includegraphics{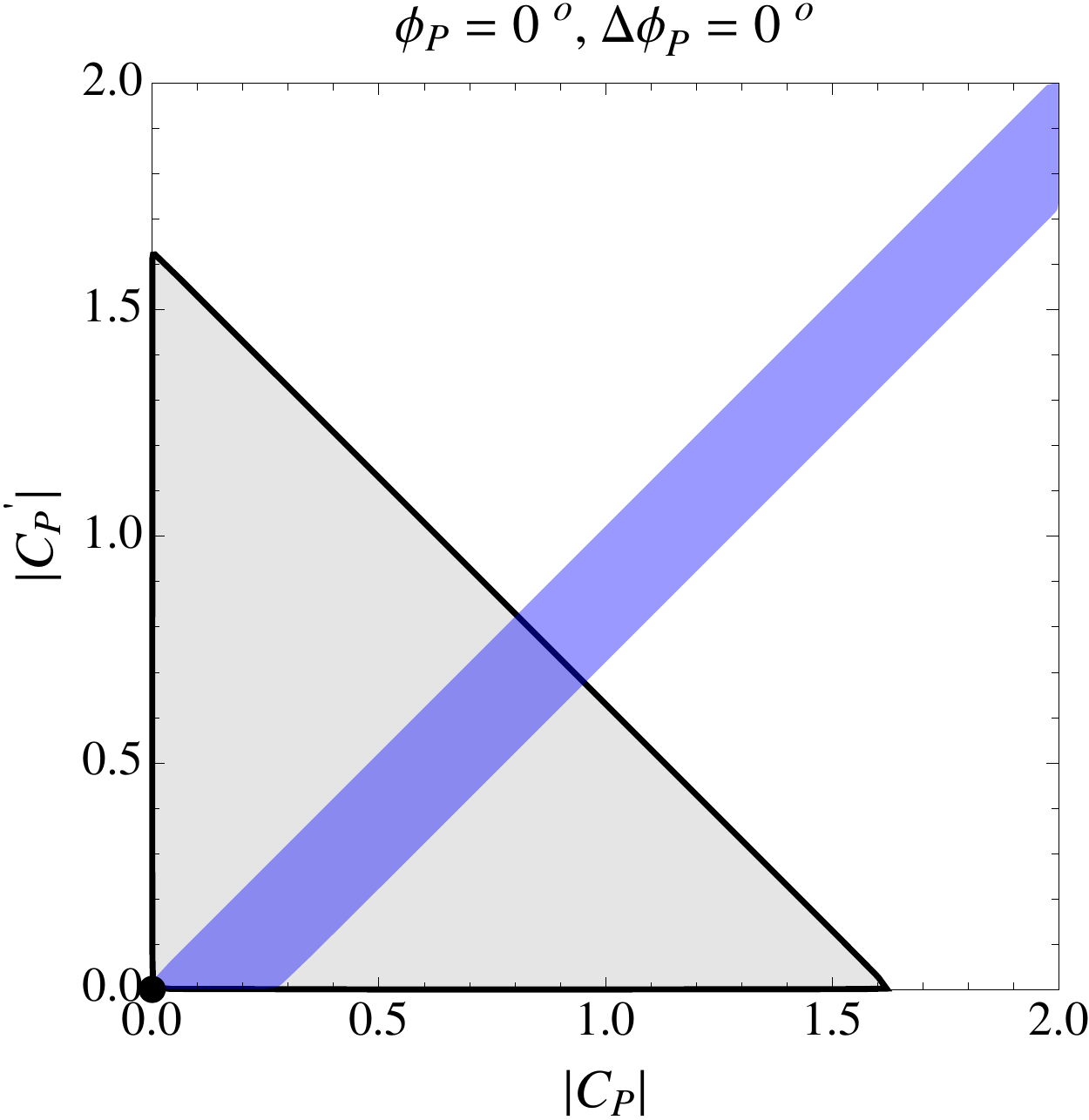}}} 
{\resizebox{5.2cm}{!}{\includegraphics{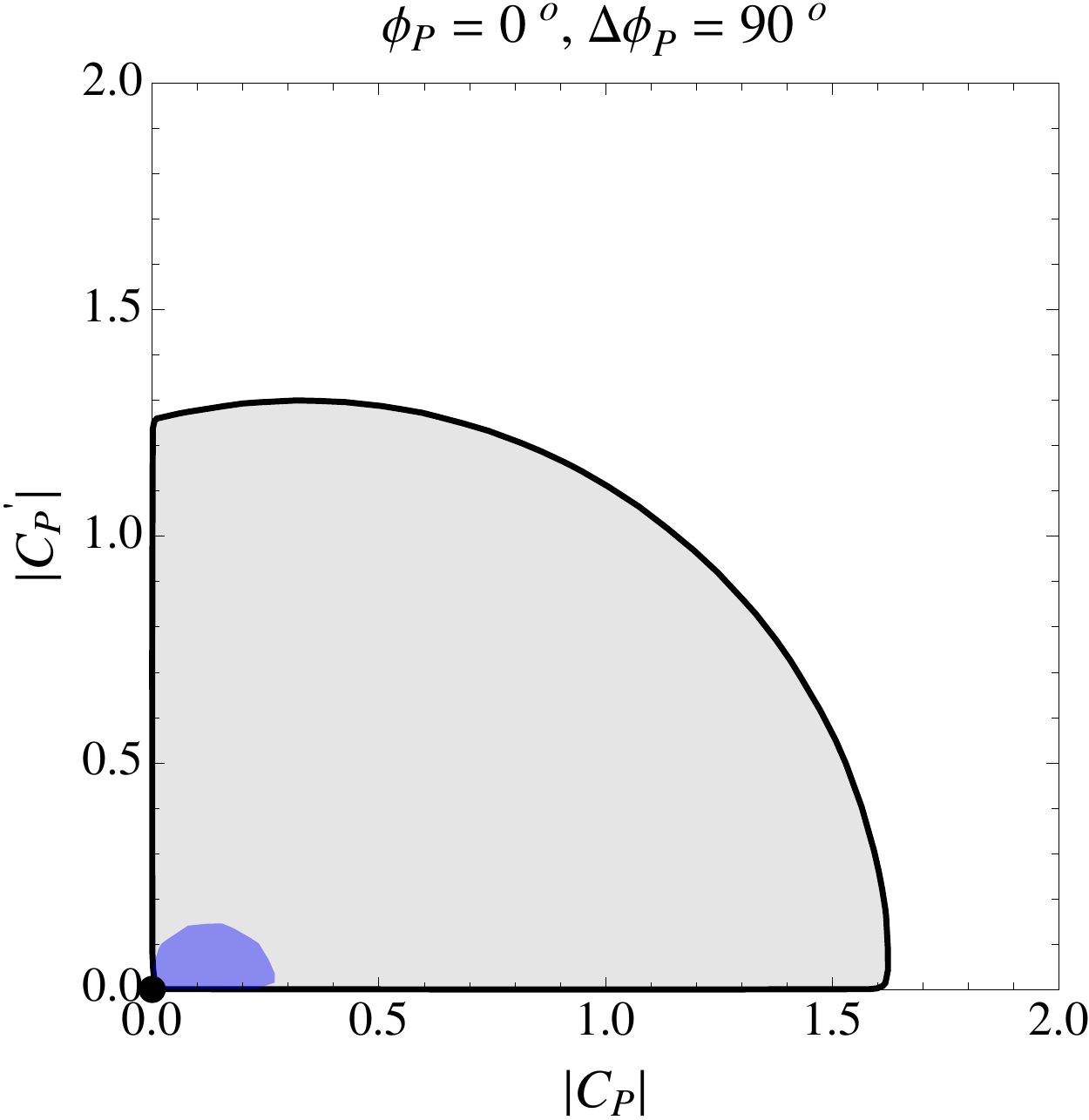}}} 
{\resizebox{5.2cm}{!}{\includegraphics{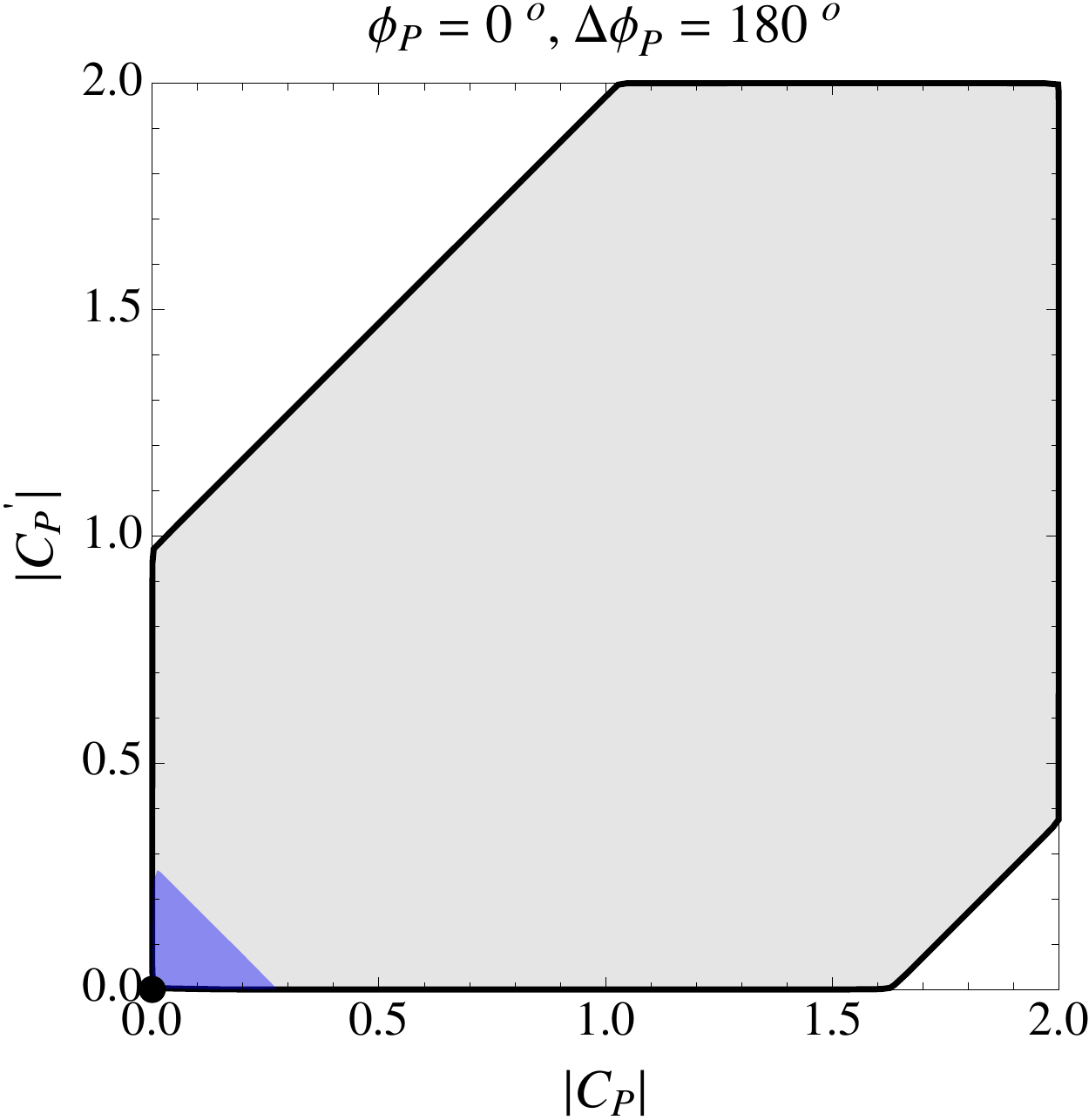}}} 
\caption{\label{fig:7}\footnotesize{\sl 
Allowed values for $ \left| C_{P} \right|$  and $ \left|  C_{P}^{\prime} \right|$ correspond to the overlap of the constraints obtained from ${\rm Br} \left(B \to K\ell^+ \ell^- \right)$ (light shaded area) and  ${\rm Br} \left(B_s \to \mu^+ \mu^- \right)$ (dark shaded area). Illustration is provided for three various values of the relative phase $\Delta \phi_P$ and for $\phi_P=0$. }} 
\end{center}
\end{figure}

\begin{figure}[t!]
\begin{center}
{\resizebox{13cm}{!}{\includegraphics{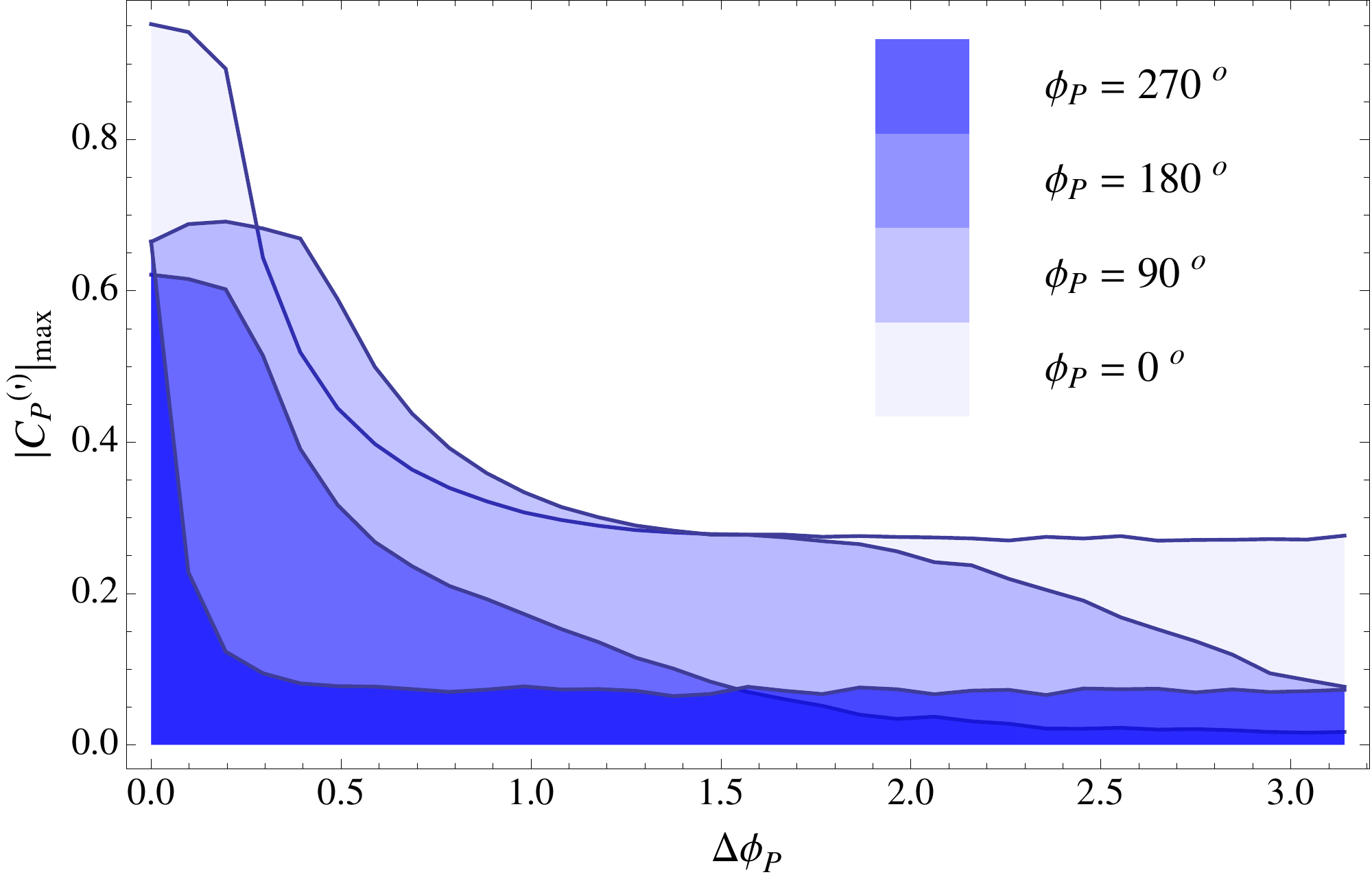}}} 
\caption{\label{fig:8}\footnotesize{\sl 
Available values of $ \left| C_{P} \right|$  or $ \left|  C_{P}^{\prime} \right|$ allowed by the experimental results~(\ref{bsmumu-exp}) and~(\ref{Kll-3}), are plotted as a function of the relative phase $\Delta \phi_P=\phi_P^\prime - \phi_P$, for four different values of $\phi_P$ specified in the legend.}} 
\end{center}
\end{figure}

We first fixed $\phi_P=0$ and varied the relative phase $\Delta \phi_P\in [0,\pi]$. Typical examples are shown in fig.~\ref{fig:7}.
We then let $\phi_P\in (0, 2\pi]$, and observe that the possible values of $ \left| C_{P} \right|$  and $ \left|  C_{P}^{\prime} \right|$ consistent with~(\ref{bsmumu-exp}) and~(\ref{Kll-3}) are smaller than in the $\phi_P=0$ case. For any fixed $\phi_P$ the situation is similar to what we observed in the case of $ \left| C_{S} \right|$  and $ \left|  C_{S}^{\prime} \right|$  (c.f. fig.~\ref{fig:6}), namely that for larger $\Delta \phi_P$ the possible values of $ \left| C_{P} \right|$  and  $ \left|  C_{P}^{\prime} \right|$ are smaller than in the case $\Delta \phi_P=0$. In other words the most space available for NP occurs when the phases of  $ \left| C_{P} \right|$  and  $ \left|  C_{P}^{\prime} \right|$ are aligned ($\Delta \phi_P\approx 0$), and even more when the new physics phase $\phi_P\approx 0$.  
This is illustrated in fig.~\ref{fig:8}.

Furthermore the observations similar to those we made in the end of previous subsection apply also in this case:
\begin{itemize}
\item Our result that $| C_{P}^{(\prime)}| \lesssim 1.0$ for any value of  $\phi_P$ and for any  $\Delta \phi_P$, is consistent with the observed branching fraction of the inclusive decay rate~(\ref{bsll-exp}) which is modified by the presence of the pseudoscalar operator in the same way it was in the case of the scalars~\cite{Fukae:1998qy}, namely, 
\bea
 {d {\rm Br} \left(B \to X_s \mu^+ \mu^- \right) \over dq^2}\Big|_{C_P^{(\prime)}} = {3 B_0\over  2 m_b^2} \left( 1 - {q^2\over  m_b^2}\right)^2  {q^2\over m_b^2}\, \left( | C_P|^2 + |C_P^{\prime}|^2 \right)  \,.
 \eea
\item Notice also that the non-zero values of $C_{P}^{(\prime)} \neq 0$ cannot modify the SM predictions of the low-$q^2$ shapes of three transverse asymmetries,  $A_T^{(2,{\rm im},{\rm re})}$, currently studied in the  $B\to K^\ast \ell^+\ell^-$ decay at LHCb.
\end{itemize}

\subsection{Peculiar case of $C_{S,P}\neq 0$}

\begin{figure}[h!]
\begin{center}
{\resizebox{5.2cm}{!}{\includegraphics{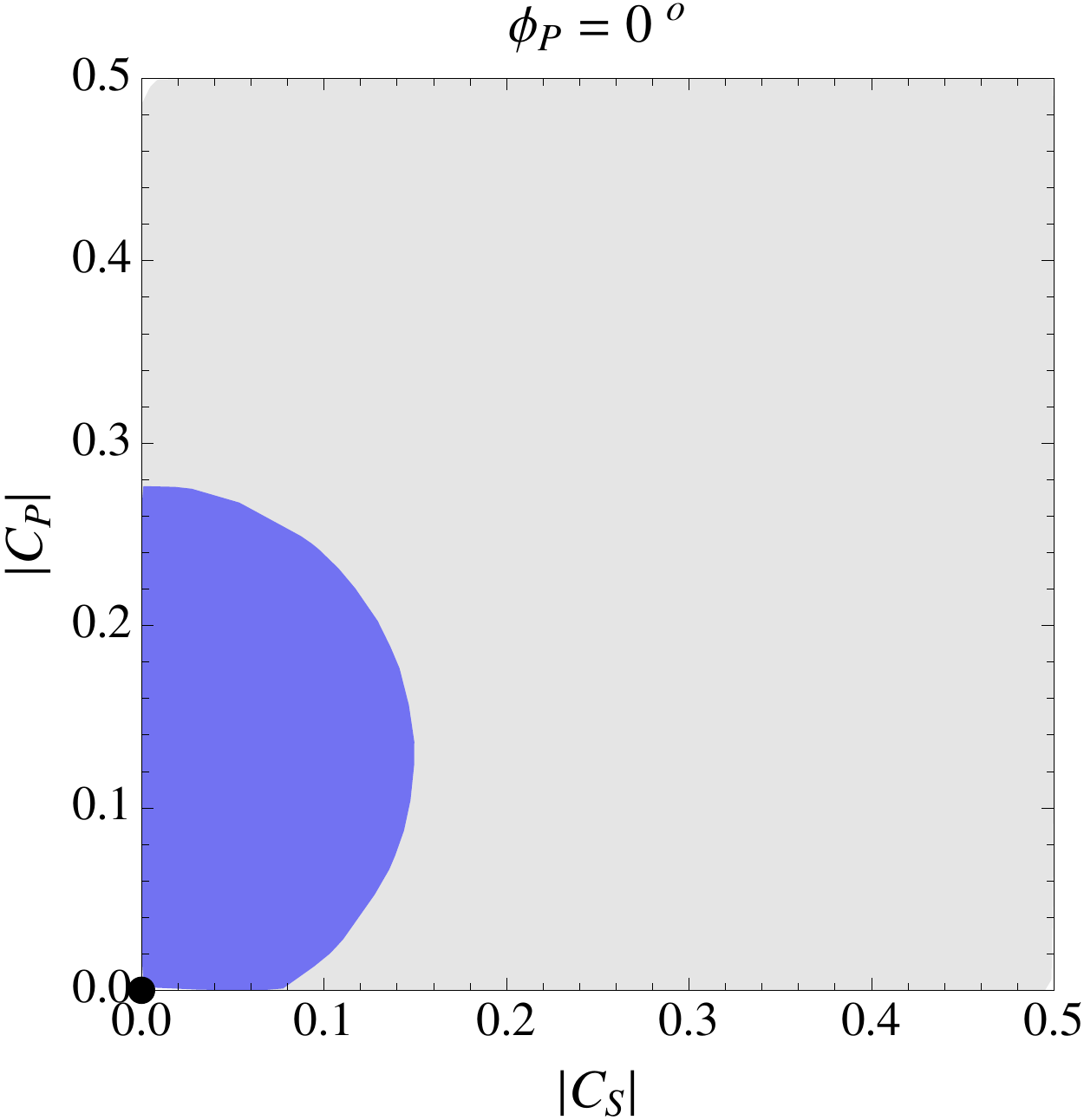}}} 
{\resizebox{5.2cm}{!}{\includegraphics{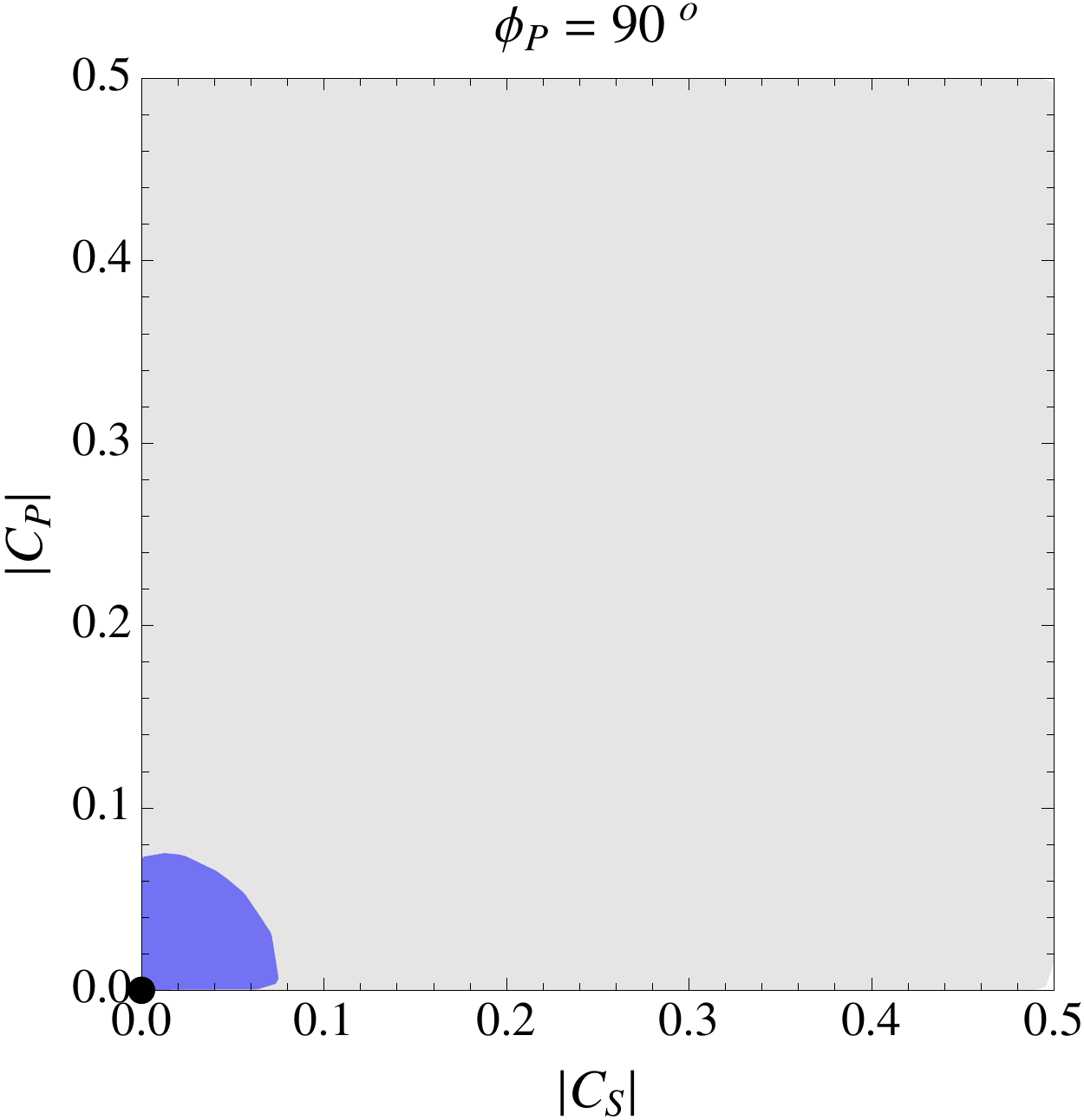}}} 
{\resizebox{5.2cm}{!}{\includegraphics{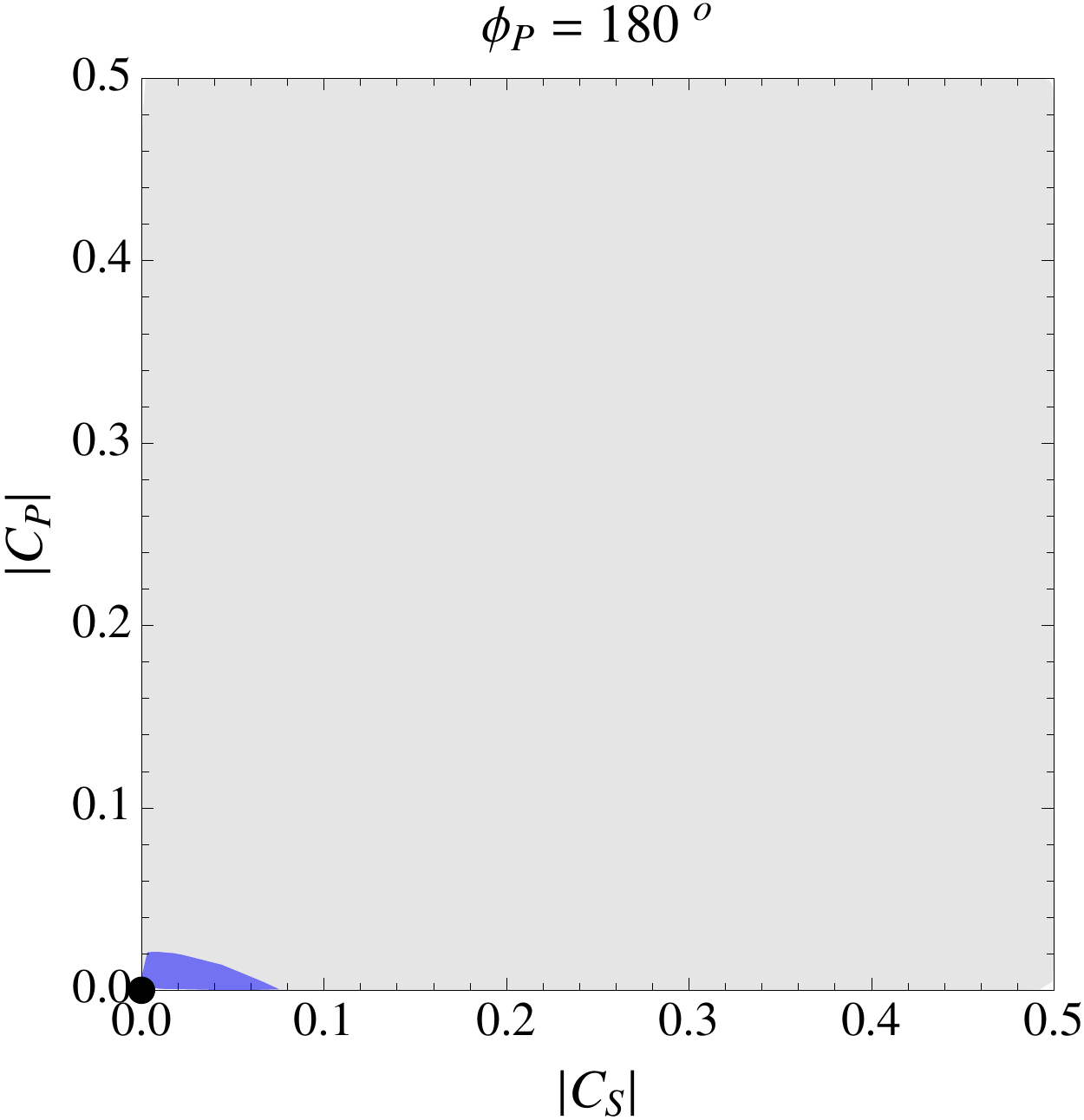}}} 
\caption{\label{fig:A}\footnotesize{\sl 
Allowed values for $ \left| C_{S} \right|$  and $ \left|  C_{P} \right|$ obtained by combining  the experimental information on  ${\rm Br} \left(B \to K\ell^+ \ell^- \right)$ (light shaded area) and the upper bound on  ${\rm Br} \left(B_s \to \mu^+ \mu^- \right)$ (dark shaded area). Illustration is provided for three various values of the relative phase $\phi_P$. }} 
\end{center}
\end{figure}

\begin{figure}[h!]
\begin{center}
{\resizebox{5.2cm}{!}{\includegraphics{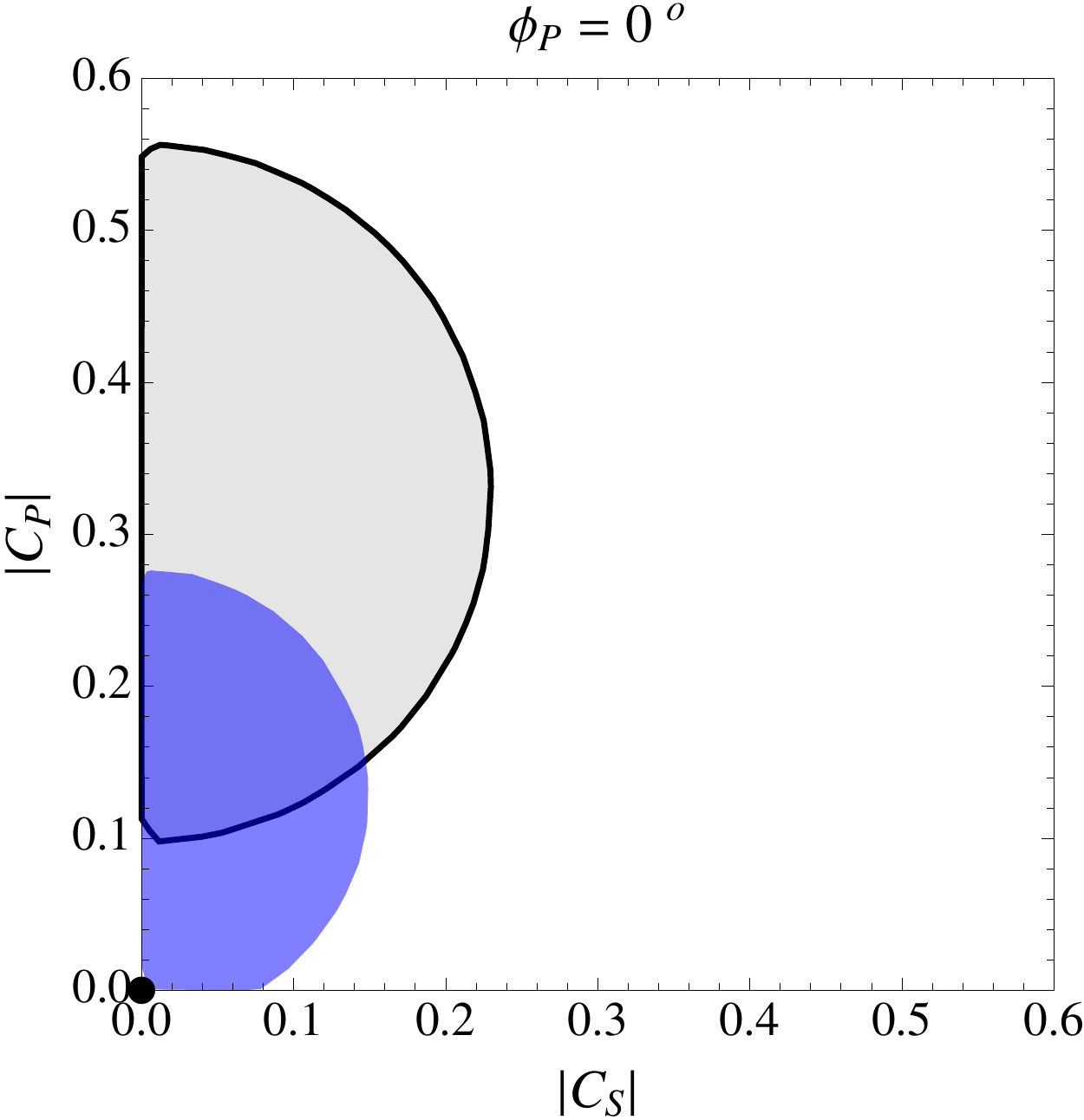}}} 
{\resizebox{5.2cm}{!}{\includegraphics{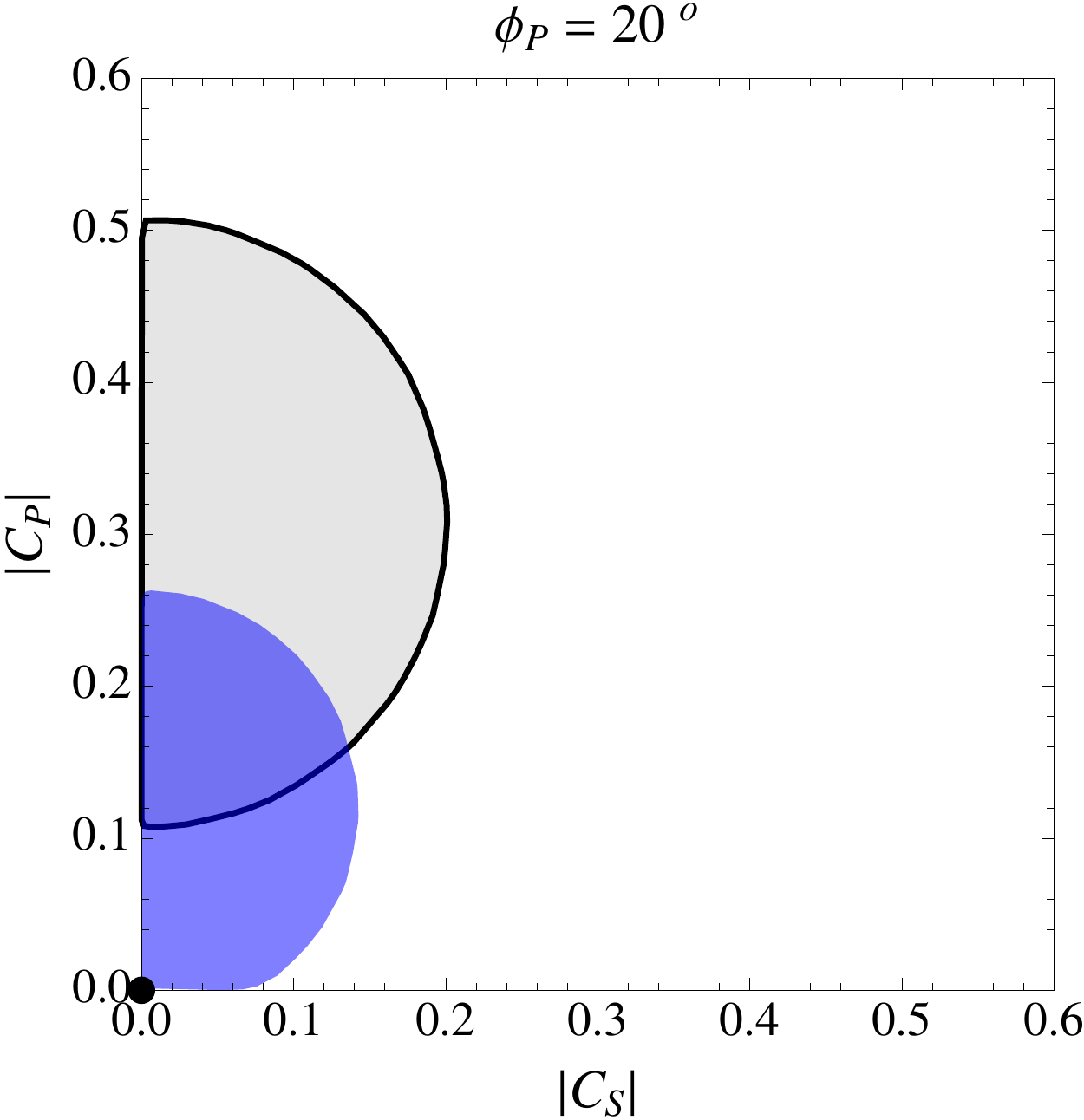}}} 
{\resizebox{5.2cm}{!}{\includegraphics{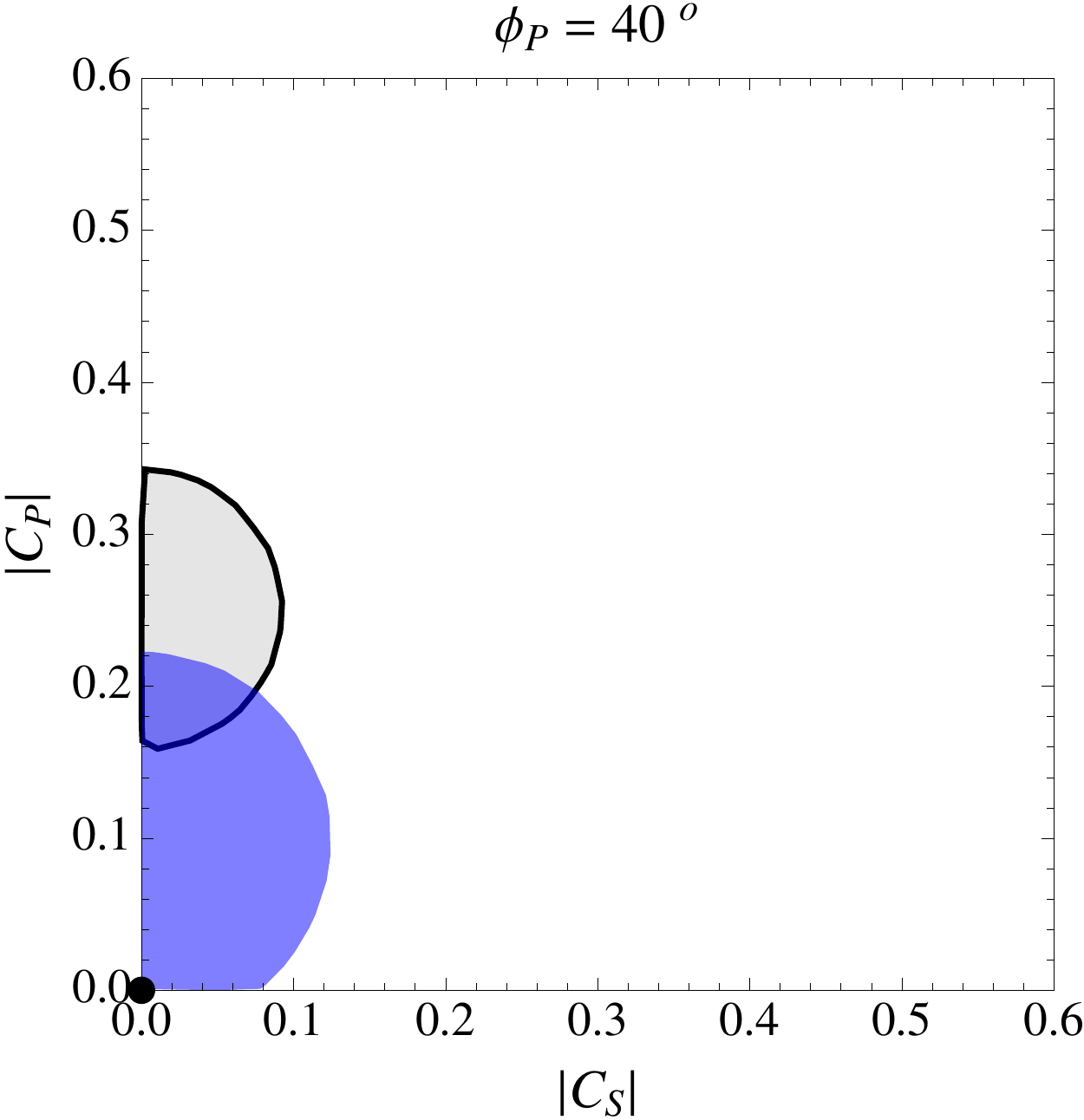}}} 
\caption{\label{fig:9}\footnotesize{\sl 
Same as in fig.~\ref{fig:A}, except that the errors on the hadronic form factors relevant to $B \to K\ell^+ \ell^- $ are reduced by $20\%$, and we plot the cases with $\phi_P=0$, $20^\circ$ and $40^\circ$. }} 
\end{center}
\end{figure}

Before closing this section, we would like to comment on the case, often discussed in the literature, in which the NP can couple via $C_{S,P}\neq 0$ but with $C_{S,P}^\prime =0$. The available range of values for $|C_{S,P}|\neq 0$ consistent with the constraints  provided by ${\rm Br} \left(B \to K\ell^+ \ell^- \right)$ and ${\rm Br} \left(B_s \to \mu^+ \mu^- \right)$  is depicted in fig.~\ref{fig:A}.
As in the previous cases the largest range of $|C_{S,P}|\neq 0$ is obtained when the pseudoscalar coupling is real, $\phi_P=0$. The result is of course invariant with respect to the change of the phase $\phi_S$. A particularly important observation that can be made in this case is that  the current constraint provided by $B\to K\ell^+\ell^-$ is redundant, but that situation could radically change if the errors on $B\to K$ form factors were significantly reduced. To illustrate that effect we keep the central values of the form factors fixed and reduce the errors by $20\%$. In that hypothetical situation the constraint coming from the measured and theoretically evaluated ${\rm Br} \left(B \to K\ell^+ \ell^- \right)$ are not compatible with the SM (within the $1\sigma$ accuracy), and therefore $B \to K\ell^+ \ell^- $ becomes an essential constraint to the values of possible $|C_{S,P}|$. The corresponding plots are presented in fig.~\ref{fig:9}, where we only show the cases for which the overlapping region (satisfied by both constraints) exists. For $\phi_P\gtrsim 40^\circ$ such a solution would not exist, which would be very valuable information about new physics.

The above example only further highlights the importance of reducing the errors on the $B \to K$ transition form factors by using the currently available lattice QCD configurations that include $N_{\rm f}=2$, $2+1$, and even $2+1+1$ dynamical quark flavors. 
We should also stress that the simultaneous experimental effort in measuring the partial decay width of $B \to K\ell^+ \ell^- $ at several moderately large and large values of $q^2$ would be highly welcome because for those momentum transfers the uncertainties of the form factors computed in LQCD are under much better control than those at low $q^2$'s. Effort in that direction made in ref.~\cite{exp-BKll} is highly welcome.

\section{Constraints on $C_{10}^{(\prime)}$\label{sec:4}}

In this section we focus on the NP contributions that might arise from the couplings to the operator ${\cal O}_{10}$ and ${\cal O}_{10}^\prime$, assuming that the Wilson coefficients $C_{S,P}^{(\prime)}=0$, as in the SM.~\footnote{Here we also tacitly assume that the $C_{7,9,T}^{(\prime)}$, which enter the expression for ${\rm Br} \left(B \to K\ell^+ \ell^- \right)$, remain at their SM values.}
Specific realizations of the two-Higgs doublet models have been discussed in great details in refs.~\cite{Chankowski:2000ng,Logan:2000iv}. Note that  our Wilson coefficients  $C_{10}^{(\prime)}$ are related to the ones defined in ref.~\cite{Chankowski:2000ng} as:
\bea
C_{10}= X\left( C_{LR}^V - C_{LL}^V \right)^\ast\,,\quad
C_{10}^\prime = X \left( C_{RR}^V - C_{RL}^V \right)^\ast\,,
\eea
where $X$ is the same one defined after eq.~(\ref{eq:chank}). 
From refs.~\cite{Chankowski:2000ng,Logan:2000iv} we learn that:
\begin{itemize}
\item The $Z^0$-penguin, with a charged Higgs running in the loop, gives rise to
\bea
C_{10}^{Z^0H^+} \propto + {m_t^2\over m_W^2}  {1\over \tan^2\beta} \,,\quad C_{10}^{\prime Z^0H^+} \propto - {m_sm_b\over m_W^2}  {\tan^2\beta} \,.
\eea
Therefore a non-zero contribution in the scenario is conceivable for either small or large $\tan\beta$, although the large $\tan\beta$ in $C_{10}^\prime$ is suppressed by the strange quark mass. 
\item The $Z^0$-penguin, with a gluino running in the loop, is only relevant at larger $\tan\beta$ and the corresponding contributions are such that $C_{10}^{Z^0\tilde g}=C_{10}^{\prime Z^0\tilde g}$.
\item Box diagrams are highly suppressed and give no interesting contributions. 
\end{itemize}
As far as the leptoquark models are concerned, $C_{10}$ and $C_{10}^\prime$ can be non-zero in both classes of models, namely with scalar of vector leptoquarks. Interestingly, however, the change in $C_{10}^{(\prime)}$ implies the change in $C_{9}^{(\prime)}$ too. For more details about this issue see ref.~\cite{nejc}.

\begin{figure}[h!]
\begin{center}
{\resizebox{5.2cm}{!}{\includegraphics{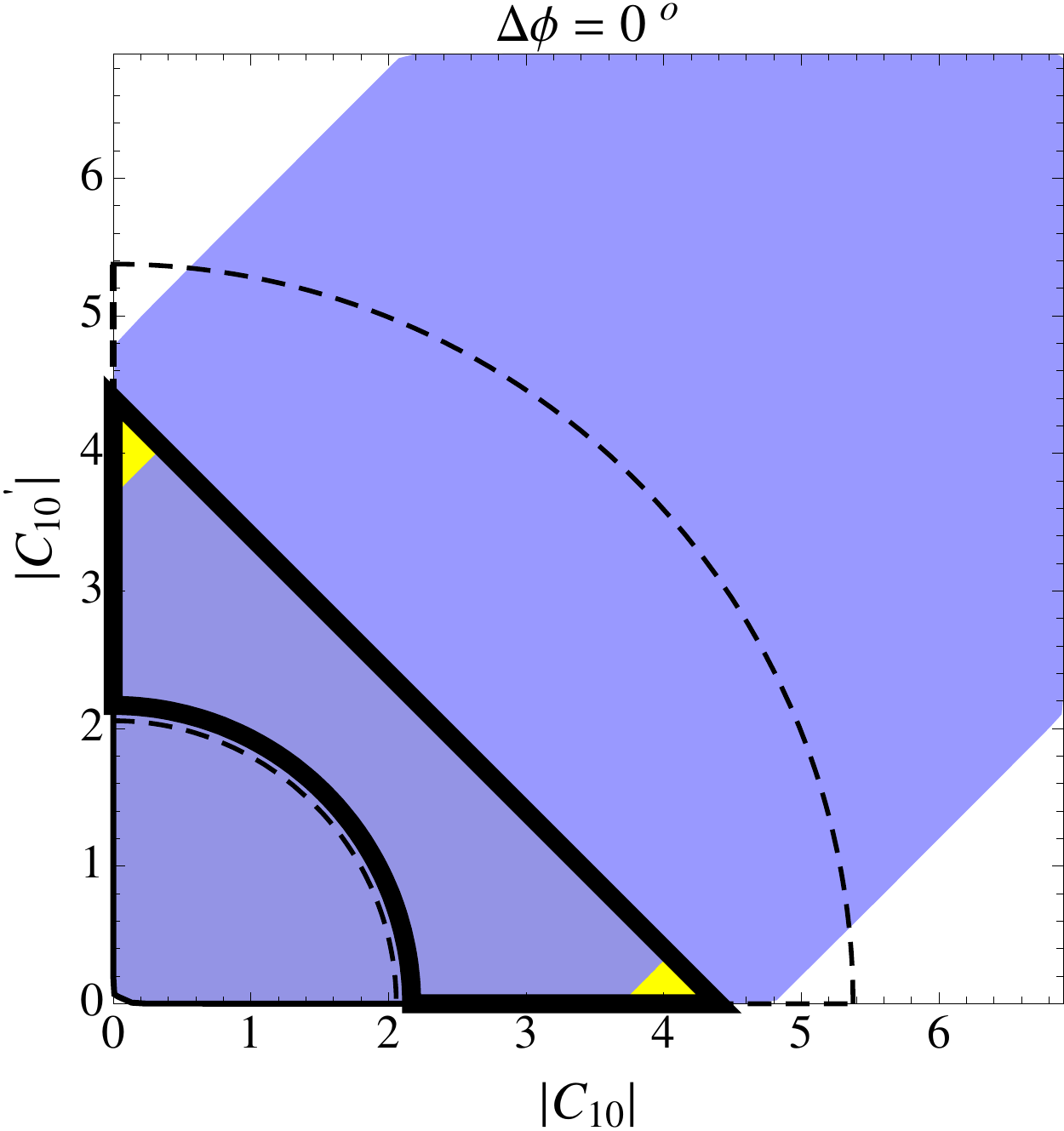}}} 
{\resizebox{5.2cm}{!}{\includegraphics{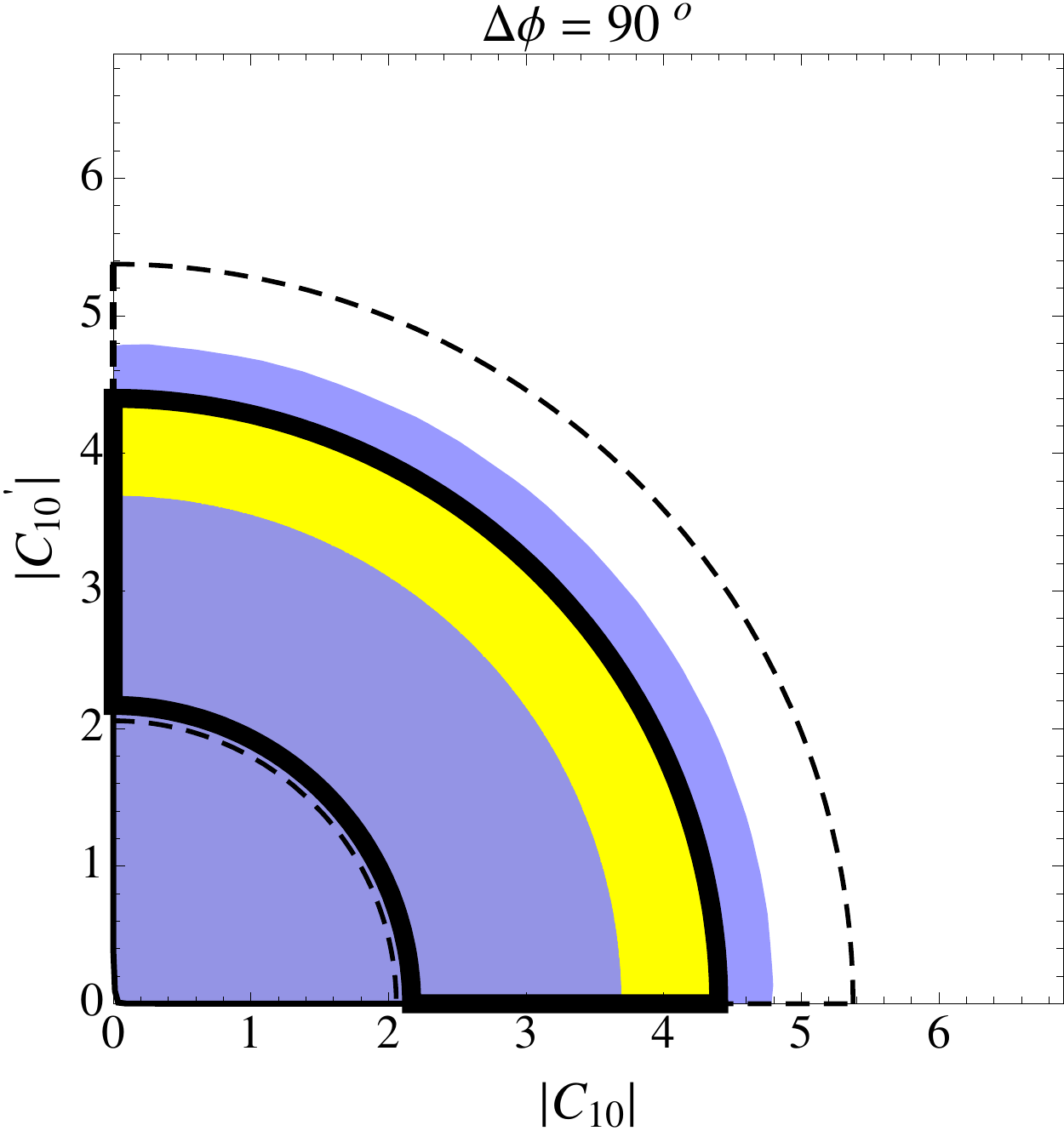}}} 
{\resizebox{5.2cm}{!}{\includegraphics{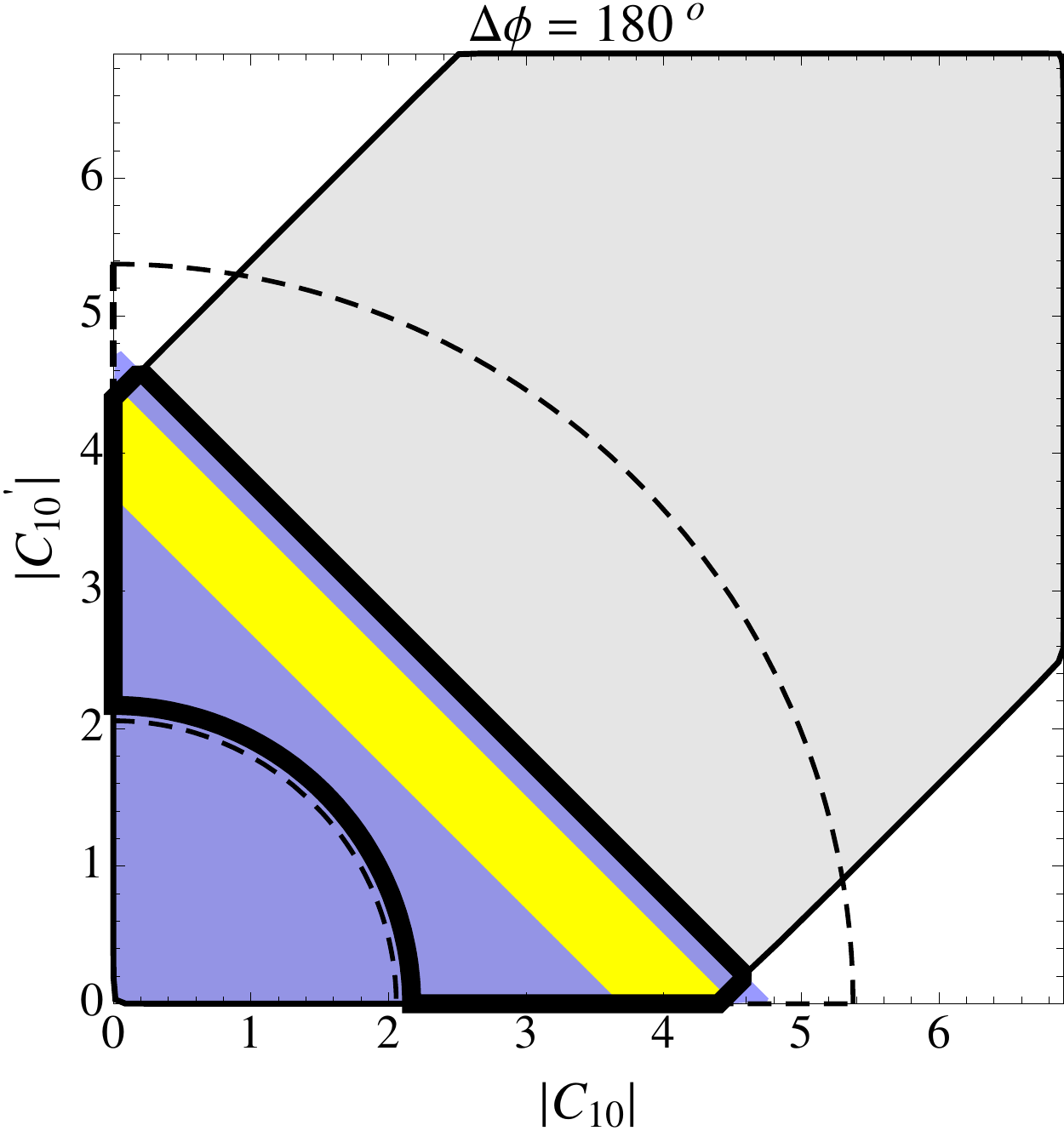}}} 
\caption{\label{fig:10}\footnotesize{\sl 
Constraint on $ \left| C_{10} \right|$  and $ \left|  C_{10}^{\prime} \right|$ obtained by combining the experimental information on ${\rm Br} \left(B \to K\ell^+ \ell^- \right)$  (light shaded area) and ${\rm Br} \left(B_s \to \mu^+ \mu^- \right)$ (dark shaded in the plots). We used the hadronic form factors and the decay constant given in Appendix~A and B. Three plots correspond to three specific choices of the relative phase $\Delta \phi = \phi_{10^\prime}-\phi_{10}$ indicated in each plot. The domain inside the dashed curve is allowed by the inclusive decay, as indicated in eq.~(\ref{constraint-inclusive}). The region satisfying all three constraints is within the thick curve. See text for the explanation about the yellow region.}} 
\end{center}
\end{figure}

In what follows we proceed in a way similar to the previous section, and use eqs.~(\ref{bsmumu-exp}) and (\ref{Kll-3}) to obtain 
\bea
\left| C_{10} + C_{10}^{\prime  } \right| \leq 4.4\,,\qquad  \left| C_{10} - C_{10}^{\prime  } \right| \leq 4.8 \qquad (1\ \sigma)\,.
\eea
It is now sufficient to study the impact of the relative phase, $\Delta \phi =\phi_{10^\prime}-\phi_{10}$, by using 
\bea\label{decomp}
&&\left| C_{10} \pm C_{10}^{\prime  } \right|^2 =  \left| C_{10}\right|^2 + \left| C_{10}^{\prime  } \right|^2\pm 2  \left| C_{10}\right|  \left| C_{10}^{\prime  } \right| \cos(\Delta \phi)\,.
\eea
In fig.~\ref{fig:10} we illustrate the resulting constraints for three distinctive cases $\Delta\phi = 0, \pi/2$, and $\pi$. 
We see that for $\Delta\phi = 0$ the main constraint comes from $B \to K\ell^+ \ell^-$, whereas in the case of  $\Delta\phi =\pi$ the decisive constraint is $B_s \to \mu^+ \mu^-$, which is easy to understand from eq.~(\ref{decomp}). In the intermediate case of $\Delta\phi = \pi/2$ the two constraints are equivalent. 

We also checked the hypothetical scenario in which the measured ${\rm Br}(B_s\to \mu^+\mu^-)$ coincides with its value predicted in the SM~(\ref{bsmumu-exp}). As a result the allowed region in the $\left| C_{10}\right|$-$\left| C_{10}^\prime\right|$ plane is depicted by the yellow stripe in fig.~\ref{fig:10}.  

\begin{figure}[h!]
\begin{center}
{\resizebox{8cm}{!}{\includegraphics{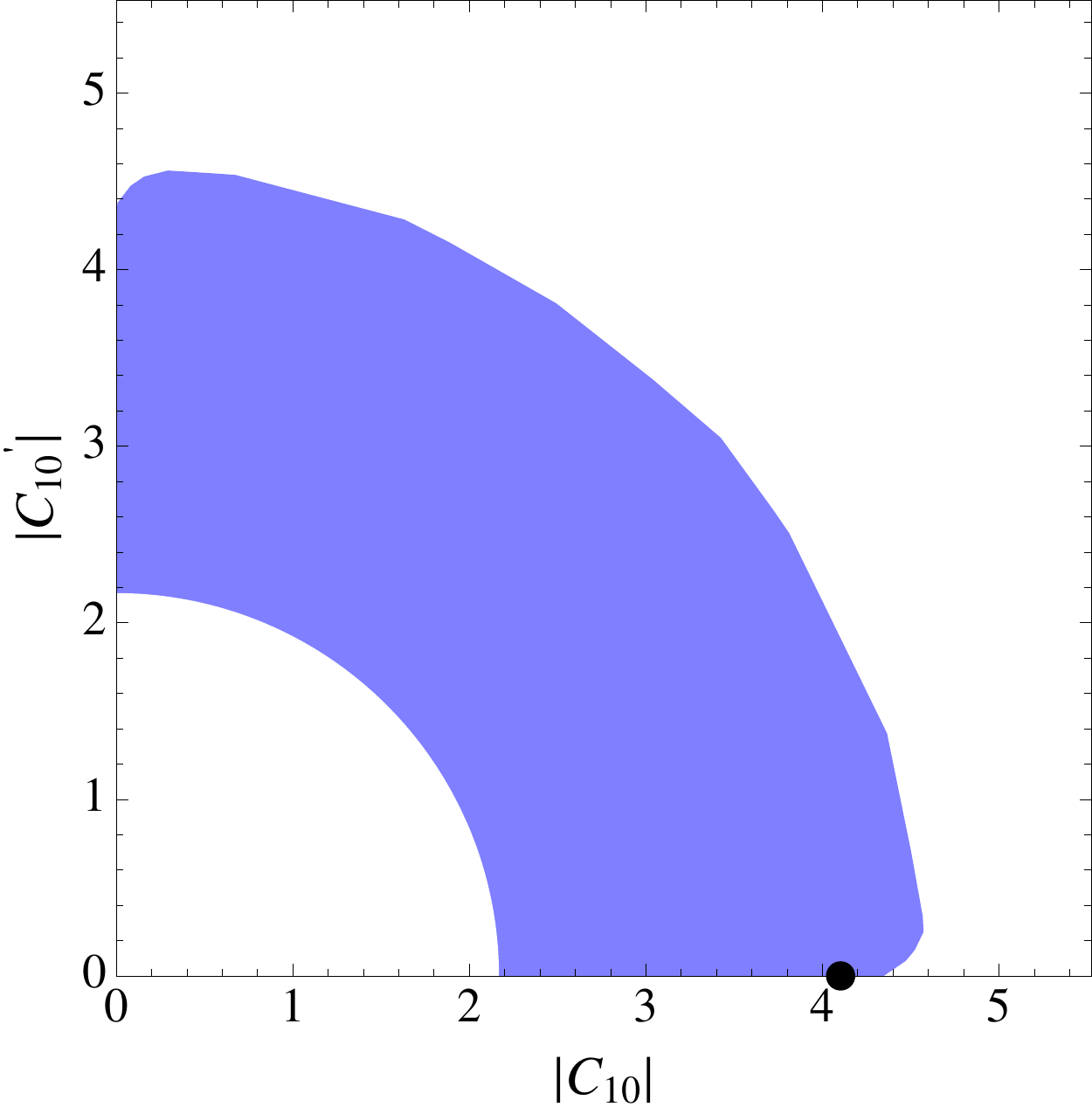}}} 
\caption{\label{fig:11}\footnotesize{\sl 
Dark region describing the possible values of $ \left| C_{10} \right|$  and $ \left|  C_{10}^{\prime} \right|$ for any relative phase $\Delta \phi \in [0,\pi]$, obtained from the measured ${\rm Br} \left(B \to K\ell^+ \ell^- \right)$, the upper bound on ${\rm Br} \left(B_s \to \mu^+ \mu^- \right)$, and the branching fraction of the partial inclusive decay rate, c.f. eqs.(\ref{bsll-exp},\ref{eq:10prime-incl}). A thick dot corresponds to the Standard Model value, $C_{10}=-4.103$.
}} 
\end{center}
\end{figure}

Contrary to the previous section, in this case the inclusive branching fraction~(\ref{bsll-exp}) provides us with a valuable new constraint. The contribution from ${\cal O}_{10}^{(\prime)}$ to the differential decay rate can be extracted from ref.~\cite{Fukae:1998qy}, and for the massless lepton pair it reads, 
 \bea
 {d {\rm Br} \left(B \to X_s \mu^+ \mu^- \right) \over dq^2}\Big|_{C_{10}^{(\prime)}} = {B_0\over m_b^2} \left( 1 - {q^2\over  m_b^2}\right)^2 \left( 1 +2  {q^2\over  m_b^2}\right) \left( \left| C_{10}\right|^2 + \left| C_{10}^\prime \right|^2\right)\,,
 \eea
and therefore we can write, 
\bea\label{eq:10prime-incl}
 10^{6} \times \int_{1\ \gev^2}^{6\ \gev^2}{d {\rm Br} \left(B \to X_s \mu^+ \mu^- \right) \over dq^2}dq^2 = 0.69(9) + 0.058(6)    \left( \left| C_{10}\right|^2 + \left| C_{10}^\prime \right|^2\right)\,.
\eea
When combined with the experimental value~(\ref{bsll-exp}) we get
\bea\label{constraint-inclusive}
4.7\leq \left| C_{10}\right| ^2+ \left| C_{10}^\prime \right|^2 \leq 28.9 \qquad (1\ \sigma)\,.
\eea
As before, we account for the $1 \sigma$ uncertainty around the central experimental value and take the lowest/largest possible values to obtain the limits in eq.~(\ref{constraint-inclusive}) which describes a disc in the  $ \left| C_{10} \right|$-$ \left|  C_{10}^{\prime} \right|$ plane, as shown in fig.~\ref{fig:10} (dashed curves). 
The situation is now more interesting as it depends considerably on the value of the relative phase. For $\Delta \phi =0$ the constraint coming from $B\to K\ell^+\ell^-$ is overwhelming and the one inferred from $B_s\to \mu^+\mu^-$ is only marginal. For $\Delta \phi =\pi$ the two constraints exchange the roles, and the most stringent constraint comes from $B_s\to \mu^+\mu^-$. In the intermediate situation with $\Delta \phi =\pi/2$ the two constraints are equivalent and have the shapes similar to that coming from $B\to X_s\mu^+\mu^-$. 
In fig.~\ref{fig:11} we show the possible values of $| C_{10}^{(\prime ) }|$ compatible with all three constraints that and for any $\Delta \phi \in [0,\pi]$.

Another difference with respect to the (pseudo-)scalar operators discussed in the previous section is that the low $q^2$ dependence of the three transverse asymmetries, $A_T^{(2,{\rm im},{\rm re})}(q^2)$, extracted from the full angular analysis of $B\to K^\ast \ell^+\ell^-$ decay, are different from their SM shapes when the $C_{10}^{(\prime)}$ are modified by the NP contributions (see ref.~\cite{damir-elia} for details). In particular the slope of the asymmetry $A_T^{({\rm re})}(q^2)$ is highly sensitive to the value of $C_{10}$, 
\bea
\left. {\partial A_T^{({\rm re})}(q^2)\over  \partial q^2}\right|_{q^2=0} 
 = R\  \frac{ C_{10} } {2 m_b  C_7   }\,,
\eea
where $R$ is a convenient ratio of the $B\to K^\ast$ form factors.~\footnote{More specifically, $R=(V/T_1)/( m_B + m_{K^*}) \approx (A_1/T_2)/(m_B-m_{K^*})$~\cite{damir-elia}.}  

\section{On importance of measuring the forward-backward asymmetry in $B\to K\ell^+\ell^-$ decay}

The observation that the spectrum of $b\to s\ell^+\ell^-$ decays in the region of large $q^2\gtrsim 15~\gev^2$'s is not plagued by the $\bar cc$ resonances opened numerous possibilities for testing theory against experiment~\cite{largeq2}. This is the region in which a major progress in taming the hadronic uncertainties by means of LQCD is possible and therefore a more reliable extraction of physics BSM from experiment should be possible. 

One quantity that we find particularly interesting to study at large $q^2$'s is the forward-backward asymmetry $A_{FB}^\mu(q^2)$. In the SM this quantity is zero and remains as such even if the NP considerably modifies the values of the Wilson coefficients $C_{7,9,10}^{(\prime)}$.  On the other hand, if the NP gives rise to the new (non-SM) Dirac structures,  $A_{FB}^\mu(q^2)$ can quite appreciably differ from zero. To our knowledge this observation was made for the first time in ref.~\cite{Alok:2008wp}. The expression for $A_{FB}^\mu(q^2)$ used in ref.~\cite{Alok:2008wp}, however, differs from the one reported in ref.~\cite{Bobeth:2007dw}. We checked both formulas and agree with the one given in ref.~\cite{Bobeth:2007dw}. 

To get a better insight in the impact of NP on $A_{FB}^\ell(q^2)$, we expand it in powers of the lepton mass and write~\footnote{For simplicity, here we take the Wilson coefficients to be real. If they were all fully complex then the expansion in eq.~(\ref{eq:afbX}) would look as:
\begin{align}
&A_{FB}^\ell(q^2)=  { \mathcal C(q^2) \over \Gamma_\ell }   { m_B -m_K \over m_b }   \sqrt{\lambda(q^2) }  f_0(q^2)\  {\rm Re}\Bigg\{ 2  \left[ (C_S+ C_S^\prime) C_T^\ast + (C_P+C_P^\prime) C_{T5}^\ast \right] \ q^2 f_T \qq    \nn \\ 
 & +   m_\ell \Big[  \bigl( (C_S+ C_S^\prime) (C_9+C_9^\prime)^\ast \bigr) (m_B + m_K) f_+ (q^2) + 2 m_ b \ \bigl( (C_S+ C_S^\prime) (C_7+C_7^\prime)^\ast \bigr) f_T(q^2)   \Big. \nn\\
& \Big. + 2 \ \bigl( (C_{10}+C_{10}^\prime) C_{T5}^\ast \bigr) \ f_T(q^2) \Big] + {\cal O}(m_\ell^2) \Bigg\} .
\nn\end{align}
}
\begin{align}\label{eq:afbX}
 A_{FB}^\ell(q^2) &=   { \mathcal C(q^2) \over \Gamma_\ell } \  { m_B -m_K \over m_b }\   \sqrt{\lambda(q^2) } \ f_0(q^2)\ \Bigg\{ 2  \left( C_S C_T + C_P C_{T5} \right) \ q^2 f_T \qq    \nn \\ 
 & +   m_\ell \Big[ C_S \ C_9 (m_B + m_K) f_+ (q^2) + 2 m_ b \left( C_S C_7   + 2 C_{T5} C_{10} \right) f_T(q^2) \Big] + {\cal O}(m_\ell^2) \Bigg\} .
\end{align}
All the quantities in the above formula have already been defined in Sec.~\ref{sectionKll}.  Measuring $A_{FB}^\ell(q^2)$ at large $q^2$'s would be highly beneficiary for our quest for NP at low energies. A separate experimental study of $A_{FB}^e(q^2)$ and $A_{FB}^\mu(q^2)$  would  help us  discern the first term from the second in (\ref{eq:afbX}). Notice that the first term is non-zero only if the NP coupling to a tensor operator is allowed. Therefore this quantity can be used to test the assumption we made in the previous sections of this paper when discussing ${\rm Br}(B\to K\ell^+\ell^-)$, namely that the $C_{T,T5}=0$, as in the SM.

Moreover, from the inclusive branching fraction~(\ref{bsll-exp}) to which the tensor operators contribute as~\cite{Fukae:1998qy},
 \bea
 {d {\rm Br} \left(B \to X_s \mu^+ \mu^- \right) \over dq^2}\Big|_{C_{T,T5}} ={8 B_0\over m_b^2} \left( 1 - {q^2\over  m_b^2}\right)^2 \left( 2 +  {q^2\over  m_b^2}\right) \left( \left| C_{T}\right|^2 + \left| C_{T5} \right|^2\right)\,,
 \eea
one gets, 
\bea\label{eq:inclT}
\int_{1\ \gev^2}^{6\ \gev^2}{d {\rm Br} \left(B \to X_s \mu^+ \mu^- \right) \over dq^2}dq^2 = 1.59(17) \times 10^{-6} \left[1+  0.66(9)  \left( \left| C_{T}\right|^2 + \left| C_{T5} \right|^2\right)\right]\,.
\eea
Contrary to the (pseudo-)scalar case in which the factor multiplying new Wilson coefficients is very small, c.f.  eq.~(\ref{eq:inclS}),  the corresponding factor multiplying  the tensor Wilson coefficients is much larger and consequently the constraint provided by eq.~(\ref{bsll-exp}) is much stronger, 
\bea\label{constrT}
 \left| C_{T}\right|^2 + \left| C_{T5} \right|^2 \leq 2.6\,.
\eea
\begin{figure}[t!]
\begin{center}
{\resizebox{10.8cm}{!}{\includegraphics{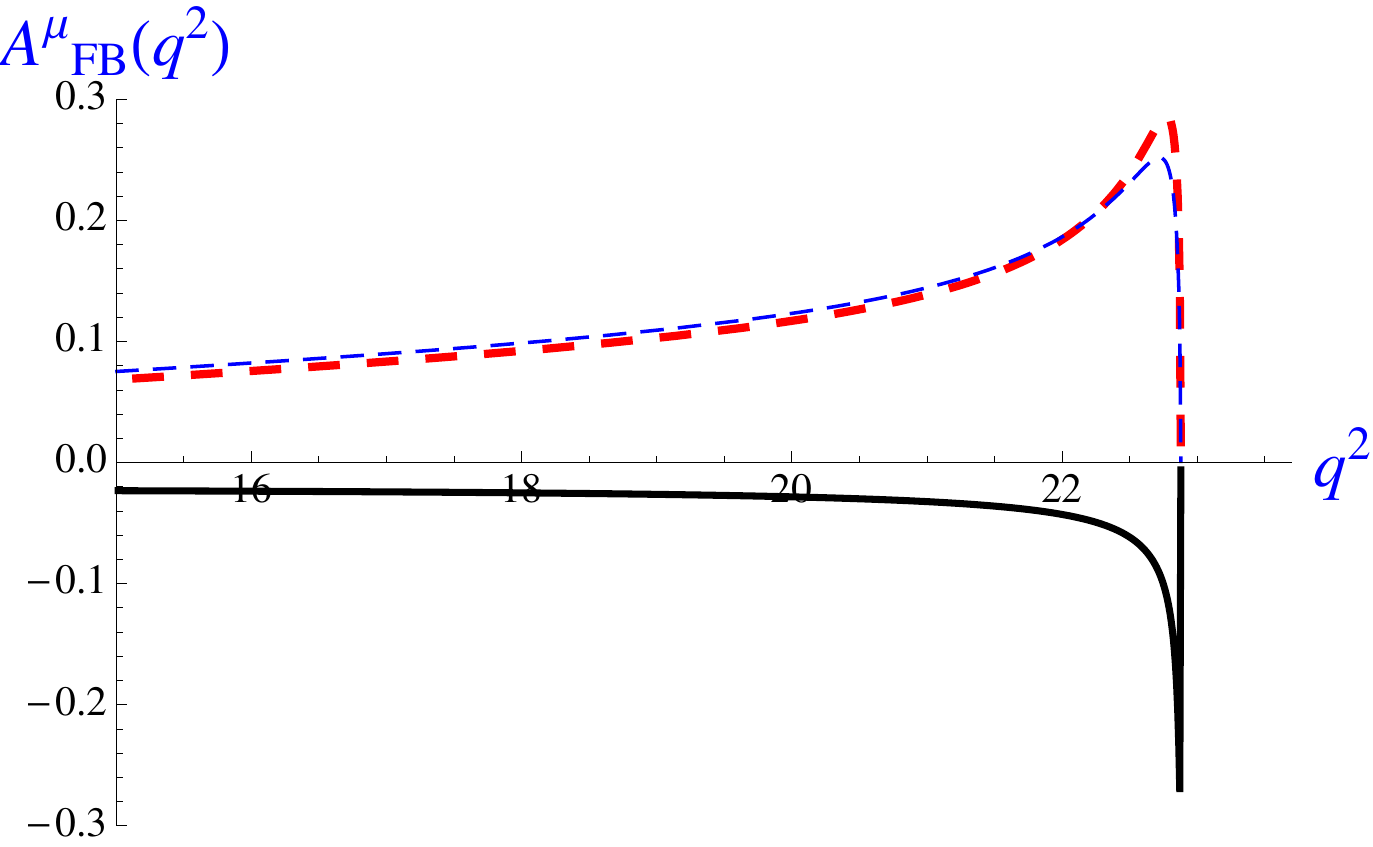}}} 
\caption{\label{fig:13}\footnotesize{\sl 
Forward-backward asymmetry in $B\to K\mu^+\mu^-$ decay. The full curve is obtained with $C_T=C_{T5}=1.1$ [consistent with eq.~(\ref{constrT})] and by keeping $C_{S,P}=0$. The dashed curves, instead, are obtained with the same $C_T=C_{T5}=1.1$, but with $C_S=0$, $C_P=1$ (thick dashed curve) or $C_P=1$, $C_S=0$ (thin dashed curve).}} 
\end{center}
\end{figure}
For the sake of illustration we  take $C_{T}=C_{T5}= 1.1$ and plot $A_{FB}^\mu(q^2)$ in fig.~\ref{fig:13} at large $q^2$'s by taking $C_{S,P}=0$. With that choice, and due to the fact that we consider the decay to the pair of muons, the $A_{FB}^\mu(q^2)\neq 0$ everywhere and is strongly enhanced near $q^2_{\rm max}=(m_B-m_K)^2$. We then switched $C_S$ or $C_P$, to illustrate the case when the first term in eq.~(\ref{eq:afbX}) is non-zero. If indeed realized in nature, this latter situation would be relatively easy to check experimentally.

\section{Summary}
In this paper we showed that  $B_s\to \mu^+\mu^-$ and $B\to
K\ell^+\ell^-$, the two actively studied decays in the $B$-physics
experiments, provide us with complementary information about the
potential NP contributions. While the decay amplitude for $B_s\to
\mu^+\mu^-$ is proportional to the difference between the Wilson
coefficients of the operators of opposite chirality, the $B\to
K\ell^+\ell^-$ involves the sum of these Wilson coefficients.

We checked the situations in which the NP enters either via the
scalar, pseudoscalar or the semileptonic operators.  To decide which
situation is verified in Nature (if any), useful information can be
obtained from the low-$q^2$ shapes of the transverse asymmetries in
$B\to K^\ast \ell^+\ell^-$ decay. Those asymmetries are being studied
in experiments and have an important advantage that the relevant
hadronic uncertainties are small. A non-zero coupling to the scalar
and/or pseudoscalar operator would not modify the low-$q^2$ shapes of
these asymmetries. From $B_s\to \mu^+\mu^-$ and $B\to K\ell^+\ell^-$
we find the absolute bounds, \bea \vert C_S^{(\prime)}\vert \lesssim
0.7, \qquad \vert C_P^{(\prime)}\vert \lesssim 1.0, \eea that are
valid for any value of the NP phases. In fact, the values for $\vert
C_{S,P}^{(\prime)}\vert$ can get considerably reduced if the non-zero
NP phases are allowed. 

In the case the coupling to the semileptonic operators ${\cal O}_{10}^\prime$ is modified by the presence of NP particles, the transverse asymmetries in $B\to K^\ast \ell^+\ell^-$ would have peculiar shapes, different from those predicted in the SM. From our study of $B_s\to \mu^+\mu^-$ and $B\to K\ell^+\ell^-$ we obtain that 
\bea
2.2 \lesssim \sqrt{ \vert C_{10} \vert^2 + \vert C_{10}^{\prime}\vert^2 } \lesssim 4.8\,,
\eea
regardless of the value of the relative NP phase. Note that the lower bound is fixed by the experimentally measured partial decay rate of the inclusive $B\to X_s\mu^+\mu^-$ decay.

In considering $B\to K\ell^+\ell^-$ we ignored the contributions from the tensor operators. 
That assumption can also be experimentally tested by measuring the non-zero forward-backward asymmetry in $B\to K\ell^+\ell^-$ decay.

Our approach of considering a pair of Wilson coefficients at a time is
orthogonal to the global fit approach adopted in many recent works.
We have checked explicitly our results with the results of Bobeth et
al in~\cite{RECENT} where a fit to real Wilson
coefficients $C_{7,9,10}$ included also the experimental observables related to
$B \to K^* \ell^+ \ell^-$ and $B \to K^* \gamma$. Our results in the
scenario with complex $C_{10}$ and $C_{10}^\prime$ (fig.~\ref{fig:11})
agree well with their presented range of real $C_{10}$. We have also
checked our allowed regions for Wilson coefficients against the
results of the global fit presented in Altmannshofer et al in~\cite{RECENT}.

In this paper we focused on the quantities that either have small
hadronic uncertainties or those for which the hadronic uncertainties
are likely to be improved soon. This is particularly the case with the
$B\to K$ form factors, that we computed in the quenched approximation
of QCD, which will soon be improved by including the effect of light
dynamical quarks. A detailed experimental information about the
partial decay rate of $B\to K\ell^+\ell^-$ in the upper range of
$q^2$'s will become particularly useful because the results for the
form factors computed on the lattice at larger $q^2$'s are more
reliable and have smaller errors.

\vspace{1 cm}
\section*{Note Added:}
While this paper was in writing the new results for ${\rm Br}(B\to K\ell^+\ell^-)$, measured at LHCb, appeared in ref.~\cite{Aaij:2012cq}. Their value is lower than the one reported by BaBar~\cite{exp-BKll},  and the agreement with the current theoretical estimate of the same quantity is only at the $2\sigma$-level. Once the more reliable estimate of the $B\to K$ form factors computed on the lattice become available, we will repeat the analysis presented here by including the $1$, $2$, $3 \sigma$ effects.

\section*{Acknowledgments}
We would like to thank S.~Descotes-G\'enon and A.~Tayduganov for discussions. Comments by D.~Straub and S.~Stone are kindly acknowledged too. 
Research by N.K. has been supported by {\sl Agence Nationale de la Recherche}, contract LFV-CPV-LHC ANR-NT09-508531. F.M. acknowledges financial support from FPA2010-20807 and the Consolider
CPAN project.

\newpage
\section*{Appendix A: $B\to K$ Form Factors} 
The results for the form factors used in this paper are obtained from the analysis following the same procedure as the one explained in detail in ref.~\cite{Abada:2000ty}, but by using the (quenched) gauge field configurations obtained at finer lattice spacing [$a^{-1}=3.8(1)$~GeV]. Those configurations have been used to compute the $B\to K^\ast \gamma$ form factors in ref.~\cite{Becirevic:2006nm} and we refer the reader to that paper for lattice details.

Besides the form factors $f_+(q^2)$ and $f_0(q^2)$ computed along the lines explained in ref.~\cite{Abada:2000ty}, we also computed the tensor form factor $f_T(q^2)$ appearing in eq.~(\ref{matrix-elements}). Note that the tensor density depends on the renormalization scale that we have set to $\mu=m_b$, the same scale at which the corresponding Wilson coefficients have been computed.

It is easy to extend the parameterization of ref.~\cite{Becirevic:1999kt} to include the form factor $f_T(q^2)$ and keep the minimal number of parameters needed to describe the $q^2$ dependence of all three form factors. In terms of poles exchanged in the $t$-channel,  the $q^2$-dependence of $f_T(q^2;\mu)$ is driven by the states with $J^P=1^-$.~\footnote{ 
Notice in particular that couplings to $1^+$ states are ruled out when both external states are pseudoscalars.} The lowest such a state is $B_s^\ast$, whose couplings to the vector and tensor bilinear quark operators are defined via
\bea
&&\langle 0 \vert \bar s\gamma_\mu b \vert B^\ast (p, \varepsilon_r)\rangle =  \varepsilon^{\ast r}_\mu m_{B^\ast} f^V_{B_s^\ast}\,,
  \cr
&&\cr
&&\langle 0 \vert \bar s\sigma _{\mu \nu}b \vert B^\ast (p, \varepsilon_r)\rangle =  
i  \left(p_\mu \varepsilon^{\ast r}_\nu - p_\nu \varepsilon^{\ast r}_\mu \right) f^T_{B^\ast_s}(\mu)\,.
\eea 
The nearest pole contribution then reads,
\bea
\langle  K(k)\vert \bar s\gamma_\mu b  \vert B(p)\rangle^{\rm pole} &=& \sum_r {\langle 0 \vert  \bar s\gamma_\mu b \vert B_s^\ast (q,\varepsilon_r)\rangle\ \langle B^\ast_s(\varepsilon_r)\vert B K\rangle  \over q^2 - m_{B^\ast_s}^2}\cr
 &=& -{1\over 2} \left( p^\mu+k^\mu - {m_B^2 - m_K^2 \over q^2 } q^\mu \right)  {m_{B^\ast_s} f^V_{B_s^\ast} g_{B_s^\ast B K}  \over q^2 - m_{B^\ast_s}^2}\,,\cr
 \Rightarrow f_+^{\rm pole}(q^2) &=& { -\frac{1}{2}m_{B^\ast_s} f^V_{B_s^\ast} g_{B_s^\ast B K}   \over q^2 - m_{B^\ast_s}^2}\,,
\eea
where we used the standard definition $\langle B^\ast_s( \varepsilon_r) \vert B K\rangle = g_{B_s^\ast B K}\ (k\cdot \varepsilon_r)$. 
Similarly, for the matrix element of the tensor quark operator we have
\bea
\langle  K(p_K)  \vert \bar s\sigma _{\mu \nu}b  \vert B(p_B)\rangle^{\rm pole} &=& \sum_r {\langle 0 \vert  \bar s\sigma _{\mu \nu}b \vert B_s^\ast (q,\varepsilon_r)\rangle\ \langle B^\ast_s(\varepsilon_r)\vert B K\rangle  \over q^2 - m_{B^\ast_s}^2}\,\cr
 \Rightarrow f_T^{\rm pole}(q^2;\mu) &=& { -\frac{1}{2}(m_{B}+m_K) f^T_{B_s^\ast}(\mu) g_{B_s^\ast B K}   \over q^2 - m_{B^\ast_s}^2 }\,,
\eea
and therefore the residua of these two form factors are related, i.e.
\bea\label{residua}
\underset{q^2\to m_{B^\ast_s}^2}{\mathrm{ Res}}
f_T(q^2;\mu) = { m_B + m_K\over m_{B^\ast_s}} {f^T_{B_s^\ast}(\mu)  \over f^V_{B_s^\ast} } \underset{q^2\to m_{B^\ast_s}^2}{\mathrm{ Res}} f_+(q^2) \,.
\eea
\begin{figure}[t!]
\begin{center}
{\resizebox{8cm}{!}{\includegraphics{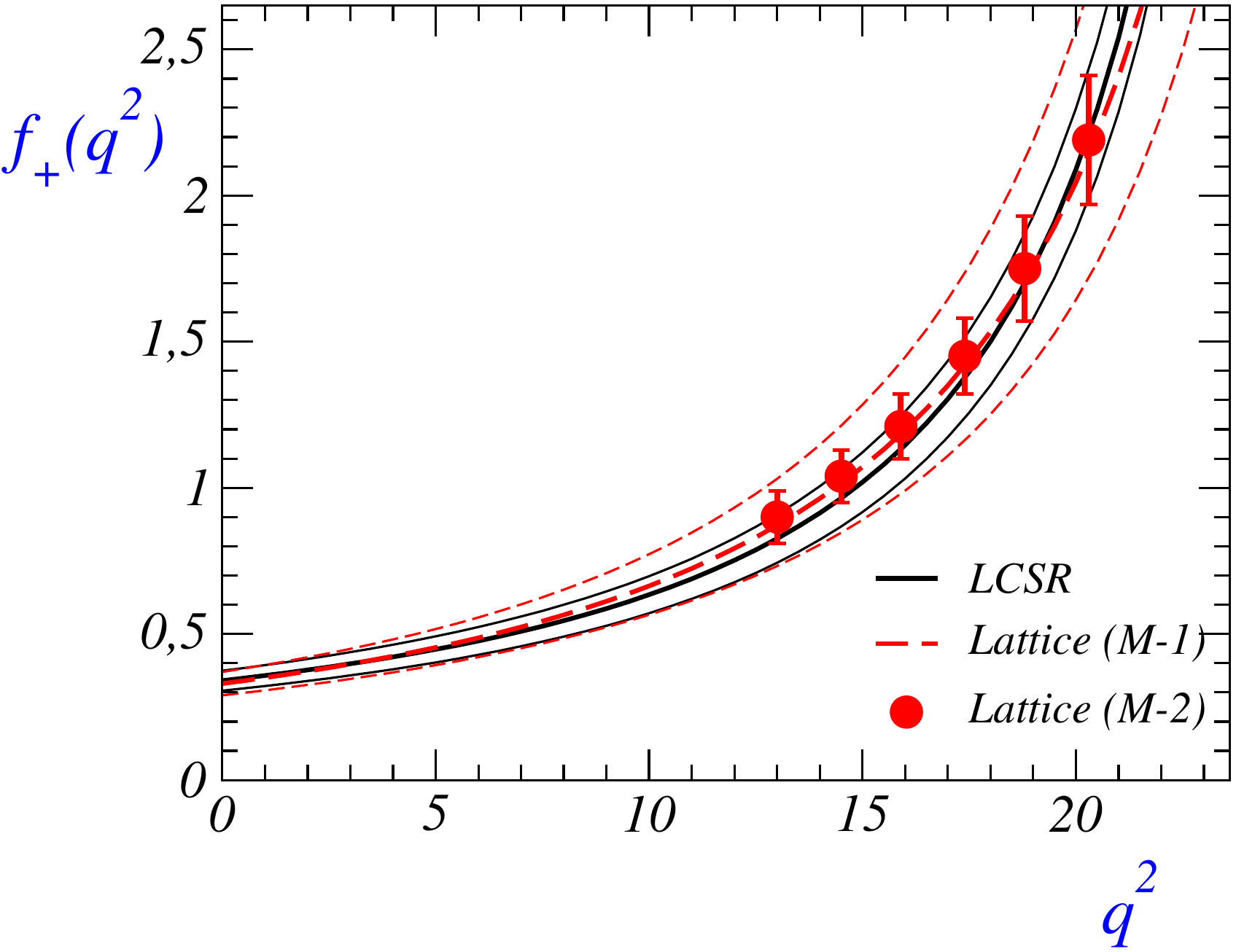}}} 
{\resizebox{8cm}{!}{\includegraphics{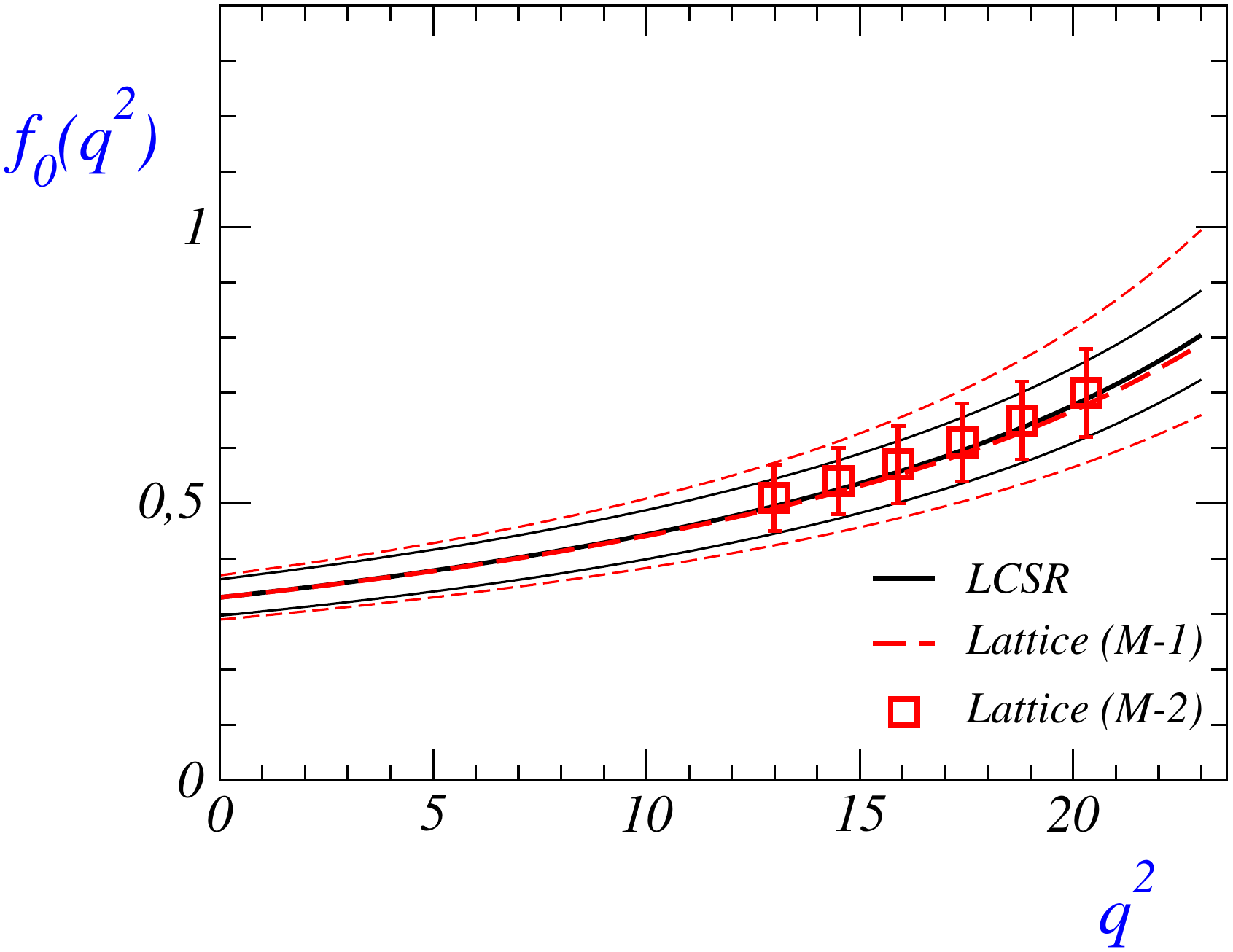}}} 
{\resizebox{8cm}{!}{\includegraphics{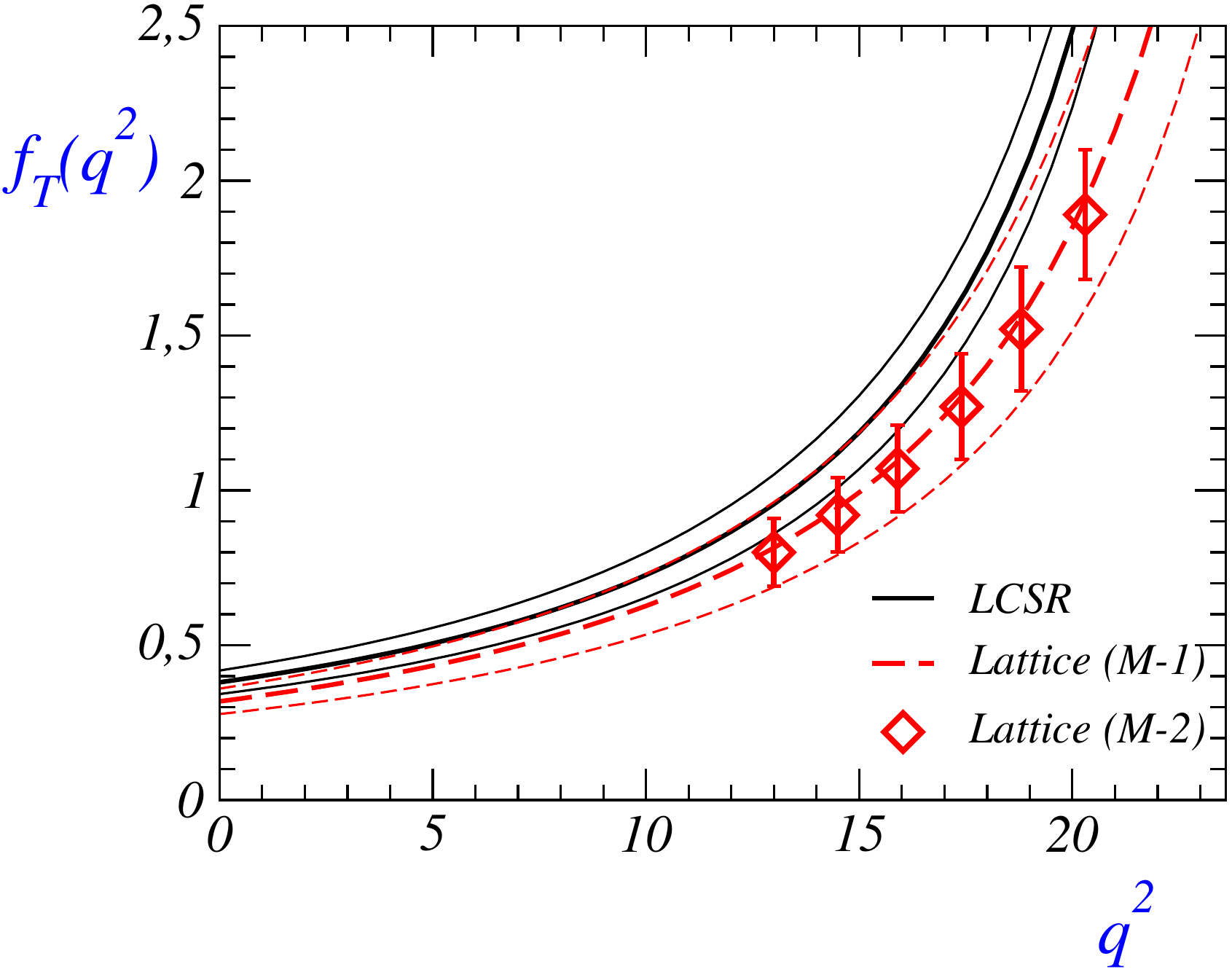}}} 
\caption{\label{fig:ffs}\footnotesize{\sl 
Comparison of the $q^2$-dependence of the $B\to K$ transition form factors as obtained in quenched lattice QCD with the predictions based on using the QCD sum rules near the light cone.
}} 
\end{center}
\end{figure}
It is quite remarkable to note that the pole dominance, that is expected to be a reasonable at large $q^2$'s (small recoils), leads to a simple proportionality relation between $f_T(q^2)$ and $f_+(q^2)$, namely
\bea
{f_T(q^2;\mu) \over f_+(q^2)} = { m_B + m_K\over m_{B^\ast_s}} {f^T_{B_s^\ast}(\mu)  \over f^V_{B_s^\ast} } \,,
\eea
which is essentially verified in the low $q^2$ region (large recoils), as obtained in ref.~\cite{jerome}:
\bea
 {f_T(q^2) \over f_+(q^2)} \approx { m_B + m_K\over m_{B}} \,.
\eea
On the basis of eq.~(\ref{residua}) it is judicious to define the scale independent 
\bea
\widetilde f_T(q^2) ={ m_{B^\ast_s}   \over m_B + m_K} {f^V_{B_s^\ast}   \over f^T_{B_s^\ast}(\mu) } f_T(q^2;\mu)\,,
\eea
which then can be easily included in the parameterization of ref.~\cite{Becirevic:1999kt} as
\bea\label{bk2}
&&f_0(q^2) = {c (1 - \alpha)\over 1 - q^2/(\beta m^2_{B^\ast_s}) }\,,\qquad f_+(q^2) = {c (1 - \alpha)\over (1 - q^2/m^2_{B^\ast_s}) (1 - \alpha q^2/m^2_{B^\ast_s}) }\,,\cr
&&\hspace*{2cm} \widetilde f_T(q^2) = {c (1 - \alpha_T )\over (1 - q^2/m^2_{B^\ast_s}) (1 - \alpha_T q^2/m^2_{B^\ast_s}) }\,,
\eea
so that only one new parameter ($\alpha_T$) is needed to fit all three form factors. The values of form factors extracted at various $q^2$ from our lattice computation are listed 
in tab.~\ref{table:ff}, where we also used $f^T_{B_s^\ast}(m_b)/f_{B_s^\ast} =0.91(3)$, computed on the same lattices.  
\begin{table}[b!]
\begin{center}
\begin{tabular}{|c|c|c|c|} 
 \hline
\hspace{-4.mm}{\phantom{\huge{l}}}\raisebox{-.2cm}{\phantom{\Huge{j}}}
 $q^2\ [{\rm GeV}^2]$  & 
 { \hspace{2mm}$f_0(q^2)$ \hspace{2mm} } & 
 { \hspace{2mm}$f_+(q^2)$ \hspace{2mm} } &
 { \hspace{2mm}$f_T(q^2)$ \hspace{2mm} } \\ 
\hline   
{\phantom{\Large{l}}}\raisebox{.2cm}{\phantom{\Large{j}}}
{ 13.0} & {${0.51(6)}$} & {${0.90(9)}$}  & {${0.80(11)}$}   \\
{\phantom{\Large{l}}}\raisebox{.2cm}{\phantom{\Large{j}}}
{ 14.5} & {${0.54(6)}$} & {${1.04(9)}$}  & {${0.92(12)}$}  \\
{\phantom{\Large{l}}}\raisebox{.2cm}{\phantom{\Large{j}}}
{ 15.9} & {${0.57(7)}$} & {${1.21(11)}$} & {${1.07(14)}$}  \\
{\phantom{\Large{l}}}\raisebox{.2cm}{\phantom{\Large{j}}}
{ 17.4} & {${0.61(7)}$} & {${1.45(13)}$}  & {${1.27(17)}$}  \\
{\phantom{\Large{l}}}\raisebox{.2cm}{\phantom{\Large{j}}}
{ 18.8} & {${0.65(7)}$} & {${1.75(18)}$}  & {${1.52(20)}$}   \\
{\phantom{\Large{l}}}\raisebox{.2cm}{\phantom{\Large{j}}}
{ 20.3} & {${0.70(8)}$} & {${2.19(22)}$}  & {${1.89(21)}$}   \\
  \hline 
\end{tabular}
\caption{\label{table:ff}{\small\sl $B\to K \ell^+\ell^-$ form factors at several values of $q^2$. The renormalization scale for the tensor form factor 
is  $\mu=m_b$. }}
\end{center}
\end{table}
In terms of the above parameters we have
\bea\label{bkparams}
&& f_+(0)=f_0(0)=0.33(4),\quad  \widetilde f_T(0) = 0.31(4)\cr
&&\hfill\cr
&&\alpha = 0.72(14),\quad  \alpha_T = 0.67(15),\quad \beta = 1.35(15)\,,
\eea
which are obtained by either extrapolating the parameters to the $B$-meson mass (M-1), or by fitting the results from tab.~\ref{table:ff} (M-2) to the parameterization~(\ref{bk2}).

We should stress that the results for the form factors presented and used in this paper are obtained in the quenched approximation of QCD and that they will be updated very soon by using the available gauge field configurations in which the effects of the light sea quarks has been included. 
We also note that the results for the form factors presented here are compatible with those obtained by using the LCSR which are parameterized as follows~\cite{ball-zwicky}: 
\bea\label{lcsr-2}
&& f_0(q^2) = \frac{0.331(4)}{1-q^2/6.12^2}\,,\qquad f_+(q^2) = \frac{0.162(21)}{1-q^2/5.41^2} +  \frac{0.173(22)}{ \left( 1 - q^2 / 5.41^2 \right)^2}\,,\nn\\
&&\hfill \nn\\
&&\hspace*{2cm} f_T(q^2) = \frac{0.161(21)}{1-q^2/5.41^2} +  \frac{0.198(25)}{ \left( 1 - q^2 / 5.41^2 \right)^2}\,,
\eea
and illustrated in fig.~\ref{fig:ffs}.

\section*{Appendix B: Numerical values of the quantities used in this work}

The values of all quantities used in this work are listed in table~\ref{tab:1}. Two comments are in order. 
\begin{itemize}
\item We use $\tau_B=\tau_{B^0}$ to respect the experimental practice when combining the charged and neutral ${\rm Br}(B\to K\ell^+\ell^-)$ decay modes. 
\item Taking the continuum results of three unquenched simulations from ref.~\cite{fBs} in the quadrature, one gets the average $f_{B_s}=0.234(6)$~GeV, which is marginally compatible with a spectacularly accurate result reported in ref.~\cite{fBsprime}. We inflated the error to include the central value of ref.~\cite{fBsprime}.
\end{itemize}
\begin{table}[h!!]
\centering 
{\scalebox{.995}{\begin{tabular}{|ccc|ccc|}  \hline \hline
{\phantom{\huge{l}}}\raisebox{-.2cm}{\phantom{\Huge{j}}}
$C_7^\eff(m_b)$& $ -0.304$ &  \cite{Bobeth:1999mk,Altmannshofer:2008dz}  &$\qquad m_\mu$&     $0.10566$~GeV &  \cite{PDG}   \\ 
{\phantom{\huge{l}}}\raisebox{-.2cm}{\phantom{\Huge{j}}}
$ C_9^\eff(m_b)$&  $4.211$ & \cite{Bobeth:1999mk,Altmannshofer:2008dz}   &  $\qquad m_b(m_b)$&     $4.25(12)$~GeV &  \cite{PDG}   \\ 
{\phantom{\huge{l}}}\raisebox{-.2cm}{\phantom{\Huge{j}}}
$C_{10}^\eff(m_b)$& $-4.103$ &  \cite{Bobeth:1999mk,Altmannshofer:2008dz}  & $\qquad m_K$ & $0.495$~GeV  & \cite{PDG}   \\  
{\phantom{\huge{l}}}\raisebox{-.2cm}{\phantom{\Huge{j}}}
$\tau_{B_s}$ & $1.47$~ps & \cite{PDG}  & $\qquad m_B$ & $5.279$~GeV  & \cite{PDG}   \\  
{\phantom{\huge{l}}}\raisebox{-.2cm}{\phantom{\Huge{j}}}
 $\tau_{B^0}$ & $1.52(1)$~ps &  \cite{PDG} & $\qquad m_{B_s}$ & $5.366(1)$~GeV  & \cite{PDG}   \\  
{\phantom{\huge{l}}}\raisebox{-.2cm}{\phantom{\Huge{j}}}
$\alpha\equiv \alpha_{\rm em}(M_Z^2) $ &  $1/128.94(3)$  & \cite{Martin:2000dd}  & $\qquad f_{B_s}$& $0.234(10)$~GeV & \cite{fBs, fBsprime}   \\  
{\phantom{\huge{l}}}\raisebox{-.2cm}{\phantom{\Huge{j}}}
$G_F$&   $1.1664\times 10^{-5}\gev^{-2}$ & \cite{PDG}  &  $\qquad |V_{tb}V_{ts}^\ast |$&   $0.0403(8)$ & \cite{Charles:2011va}  \\ 
 \hline \hline
\end{tabular} }}
{\caption{\footnotesize  \label{tab:1} All the scale dependent quantities are obtained in the $\msbar$ renormalization scheme. The Wilson coefficients are evaluated at the next-to-next-to leading logarithms. $\alpha_s(M_Z)=0.118$.}}
\end{table}

\end{document}